\documentclass[fleqn,usenatbib]{mnras}

\usepackage[T1]{fontenc}

\DeclareRobustCommand{\VAN}[3]{#2}
\let\VANthebibliography\thebibliography
\def\thebibliography{\DeclareRobustCommand{\VAN}[3]{##3}\VANthebibliography}

\usepackage{graphicx}	
\usepackage{amsmath}	
\usepackage{amssymb}

\title[Exploring variability and orbit of AB Pic b]{The ESO SupJup Survey V: Exploring Atmospheric Variability and Orbit of the Super-Jupiter AB Pictoris b with CRIRES+}

\author[Gandhi et al.]{
Siddharth Gandhi$^{1,2,3}$\thanks{E-mail: Siddharth.Gandhi@warwick.ac.uk},
Sam de Regt$^{3}$,
Ignas Snellen$^{3}$,
Paulina Palma-Bifani$^{4,5}$,
Idriss Abdoulwahab$^{4,6}$,
\newauthor{
Gaël Chauvin$^{7,4}$,
Dar\'io Gonz\'alez Picos$^{3}$,
Yapeng Zhang$^{8}$,
Rico Landman$^{3}$,
Tomas Stolker$^{3}$,}
\newauthor{
Aurora Kesseli$^{9}$,
Willeke Mulder$^{3}$,
Antoine Chomez$^{5,10}$,
Anne-Marie Lagrange$^{5,10}$,
Alice Zurlo$^{11,12}$
}
\\
$^{1}$Department of Physics, University of Warwick, Coventry CV4 7AL, UK\\
$^{2}$Centre for Exoplanets and Habitability, University of Warwick, Gibbet Hill Road, Coventry CV4 7AL, UK\\
$^{3}$Leiden Observatory, Leiden University, Postbus 9513, 2300 RA, Leiden, The Netherlands\\
$^{4}$Laboratoire Lagrange, Université Côte d’Azur, CNRS, Observatoire de la Côte d’Azur, 06304 Nice, France\\
$^{5}$LESIA, Observatoire de Paris, Univ PSL, CNRS, Sorbonne Univ, Univ de Paris, 5 place Jules Janssen, 92195 Meudon, France\\
$^{6}$Univ Grenoble Alpes, Grenoble INP - phelma, 38000 Grenoble, France\\
$^{7}$Max-Planck-Institut fur Astronomie, Konigstuhl 17, 69117 Heidelberg, Germany\\
$^{8}$Department of Astronomy, California Institute of Technology, Pasadena, CA 91125, USA\\
$^{9}$IPAC, Mail Code 100-22, Caltech, 1200 E. California Boulevard, Pasadena, CA 91125, USA\\
$^{10}$Univ Grenoble Alpes, CNRS, IPAG, F-38000 Grenoble, France\\
$^{11}$Instituto de Estudios Astrof\'isicos, Facultad de Ingenier\'ia y Ciencias, Univ Diego Portales, Av. Ej\'ercito Libertador 441, Santiago, Chile\\
$^{12}$Millennium Nucleus on Young Exoplanets and their Moons (YEMS), Chile\\
}

\date{Accepted XXX. Received YYY; in original form ZZZ}

\pubyear{2025}

\begin{document}
\label{firstpage}
\pagerange{\pageref{firstpage}--\pageref{lastpage}}
\maketitle

\begin{abstract}
A growing number of directly-imaged companions have been recently characterised, with robust constraints on carbon-to-oxygen ratios and even isotopic ratios. Many companions and isolated targets have also shown spectral variability. In this work we observed the super-Jupiter AB~Pictoris~b across four consecutive nights using VLT/CRIRES+ as part of the ESO SupJup survey, exploring how the constraints on chemical composition and temperature profile change over time using spectral line shape variations between nights. We performed atmospheric retrievals of the high-resolution observations and found broadly consistent results across all four nights, but there were differences for some parameters. We clearly detect H$_2$O, $^{12}$CO and $^{13}$CO in each night, but abundances varied by $\sim2\sigma$, which was correlated to the deep atmosphere temperature profiles. We also found differences in the $^{12}$C$/^{13}$C ratios in each night by up to $\sim3\sigma$, which seemed to be correlated with the cloud deck pressure. Our combined retrieval simultaneously analysing all nights together constrained broadly the average of each night individually, with the C/O$=0.59\pm0.01$, consistent with solar composition, and $^{12}$C$/^{13}$C~$ = 102\pm8$, slightly higher than the ISM and Solar System values. We also find a low projected rotational velocity, suggesting that AB~Pictoris~b is either intrinsically a slow rotator due to its young age or that the spin axis is observed pole-on with a $\sim90^\circ$ misalignment with its orbit inclination. Future observations will be able to further explore the variability and orbit of AB~Pictoris~b as well as for other companions. 
\end{abstract}




\section{Introduction}

\begin{figure*}
	\includegraphics[width=\textwidth]{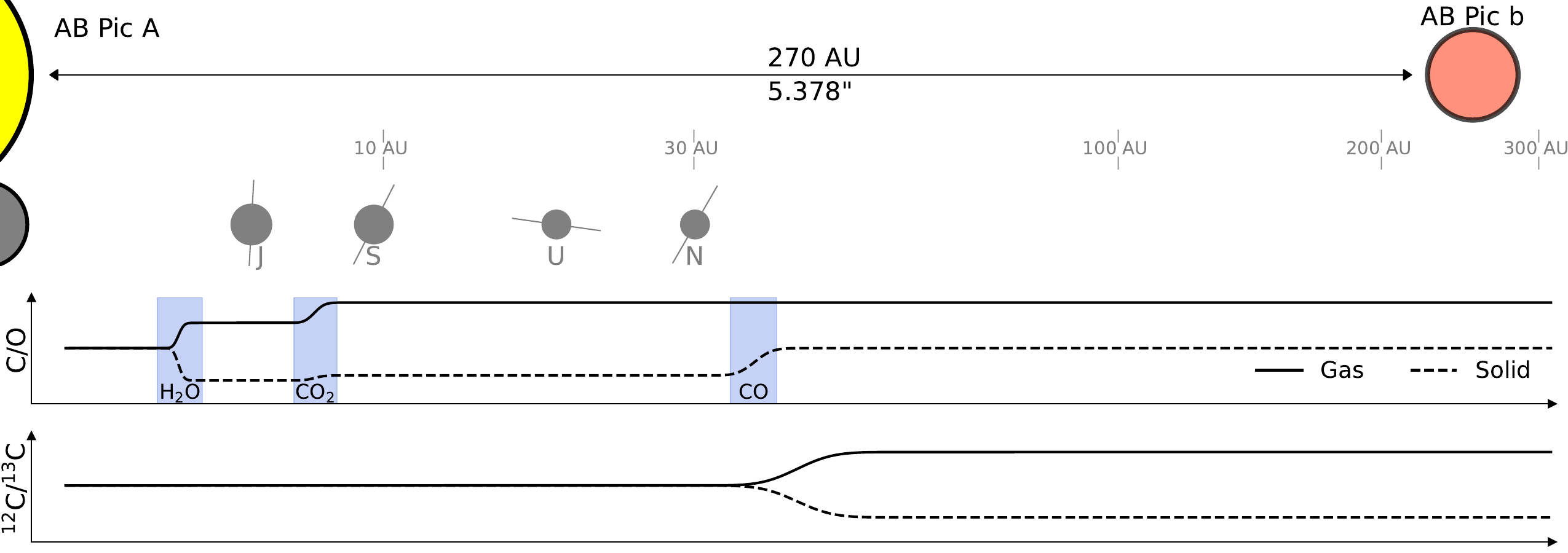}
    \caption{Schematic of the AB Pic system, showing the orbital separation of the host star to the target. We also show the orbital separations and inclinations of the Solar System gas giants and the suggested variation of the gas and solid phase C/O ratio and $^{12}$C/$^{13}$C. The blue regions show the expected locations of the snowlines for the prominent species in protoplanetary disks. Gas giant planets that form further from their host stars past the snowlines may accrete gaseous envelopes that are more C/O-rich and have higher $^{12}$C/$^{13}$C ratios.}
    \label{fig:schematic}
\end{figure*}

Directly-imaged companions at wide orbital separations represent unique targets to characterise formation at the extremes of orbital separation. They are a new frontier in exoplanet astronomy, and have been characterised in unprecedented detail thanks to developments in instrumentation and adaptive optics \citep[see e.g.][]{currie2023}. These companions have an internal source of heat from their formation and potentially deuterium burning, rather than strong incident stellar radiation as is the case for hot Jupiters. Therefore their temperature profiles are more steeply increasing with pressure, resulting in strong absorption features from spectroscopically active molecular and atomic species.

A range of chemical species have been detected in isolated objects as well as companions through retrievals, which constrain the atmospheric properties by performing Bayesian analyses and exploring a wide range of parameter space. The chemical constraints can be used to determine the carbon-to-oxygen (C/O) ratio and metallicity of these objects \citep[e.g.][]{line2015, line2017, burningham2017, lavie2017}, key parameters in planet formation \citep{oberg2011, madhu2012, mordasini2016}. Atmospheric retrievals have also been able to constrain cloud properties \citep[e.g.][]{burningham2021, vos2023} as well as chemical disequilibrium processes \citep{molliere2020, deregt2024} thanks to the often very high signal-to-noise observations. Such retrievals are now also opening up atmospheric surveys across the population of companions as well as comparative science with the compositions of close-in hot Jupiters \citep[e.g.][]{welbanks2019, xuan2024}.

A new paradigm shift in recent years has been to explore isotopic ratios \citep{molliere2019, morley2019}. The atmospheric $^{12}$C/$^{13}$C has been measured for close-in transiting planets, directly-imaged companions as well as field brown dwarfs \citep[e.g.][]{zhang2021, zhang2021_bd, line2021, costes2024, xuan2024_hip55507, deregt2024}. These measurements have pointed to a range in values, with some objects nearer $^{12}$C/$^{13}$C$\sim$30, but others with $^{12}$C/$^{13}$C values $\gtrsim$100. Thanks to the arrival of JWST, we have been able to obtain high precision constraints on the carbon isotope ratios \citep{gandhi2023_vhs, lew2024, hood2024}, and been able to explore spectroscopically weaker oxygen and nitrogen isotopes in planetary atmospheres for the first time \citep{gandhi2023_vhs, barrado2023, xuan2024_hip55507}. Recent work has also constrained the D/H ratio of a Y dwarf \citep{rowland2024}, proving the exciting potential of isotopic ratios. The isotopic ratios are influenced by several fractionation processes in protoplanetary disks, such as isotope selective photodissociation, gas/ice partitioning and isotopic exchange reactions, and are therefore a potential avenue to disentangle how companions form and migrate \citep[e.g.][]{miotello2014, zhang2021} (see Figure~\ref{fig:schematic}). Recently, \citet{landman2024} observed a directly-imaged companion, $\beta$~Pictoris~b, with the upgraded CRIRES+ (Cryogenic High-Resolution Infrared Echelle Spectrograph) on the VLT. Such facilities are ideal given their high spectral resolution allowing us to resolve many thousands of molecular lines in the spectrum, resulting in robust detections of a range of chemical species and isotopologues in the atmosphere.

Directly-imaged companions and isolated sub-stellar objects have shown significant variability over time \citep[e.g.][]{artigau2009, radigan2012, radigan2014, metchev2015, biller2015, eriksson2019, zhou2022, biller2024, liu2024}. These sub-stellar mass objects are often fast rotators \citep[e.g.][]{snellen2014, bryan2020, tannock2021, landman2024, hsu2024, morris2024} and inhomogeneities which vary across the observed face as they rotate result in spectroscopic variability. Theoretical works modelling the L/T transition have proposed that clouds, thermochemical instabilities, thermal variations and aurorae can all contribute to variability across the surface \citep{marley2012, robinson2014, hallinan2015, tremblin2016}. Three-dimensional modelling has also shown that feedback between clouds and chemistry can drive inhomogeneities in the photosphere \citep[e.g.][]{tan2021, lee2024}. Younger, lower gravity objects have shown evidence for thick clouds \citep[e.g.][]{currie2011, barman2011}, and even more variability compared with field brown dwarfs \citep{vos2019, vos2022}. Therefore, sub-stellar mass objects can often be dynamic and changing, but this gives us an excellent opportunity to explore how the physical and chemical properties vary over the planetary surface \citep[e.g.][]{crossfield2014}.

In this work we characterise the atmosphere of the directly-imaged companion AB~Pictoris~b (hereafter AB~Pic~b) through observations made with the CRIRES+ spectrograph on the VLT as part of the SupJup survey (Program ID: 1110.C-4264, PI: Snellen). AB~Pic~b appears redder on a colour-magnitude diagram than other targets from the SupJup survey of similar spectral type \citep{deregt2024}, typical for young and lower gravity objects. First discovered by \citet{chauvin2005}, this L0-1 object has a temperature of $\sim$1700~K \citep{bonnefoy2010, patience2012, bonnefoy2014, palma-bifani2023} and orbits its 13.3~Myr K1V type host star \citep{booth2021} at a projected orbital separation of 270~AU (5.378" on-sky separation), as shown in Figure~\ref{fig:schematic}. Its mass is between the range for planets and brown dwarfs and thus presents an ideal target to characterise, particularly given that we have no such analogue in the Solar System at this mass or at such a wide orbital separation.

\begin{figure*}
	\includegraphics[width=\textwidth]{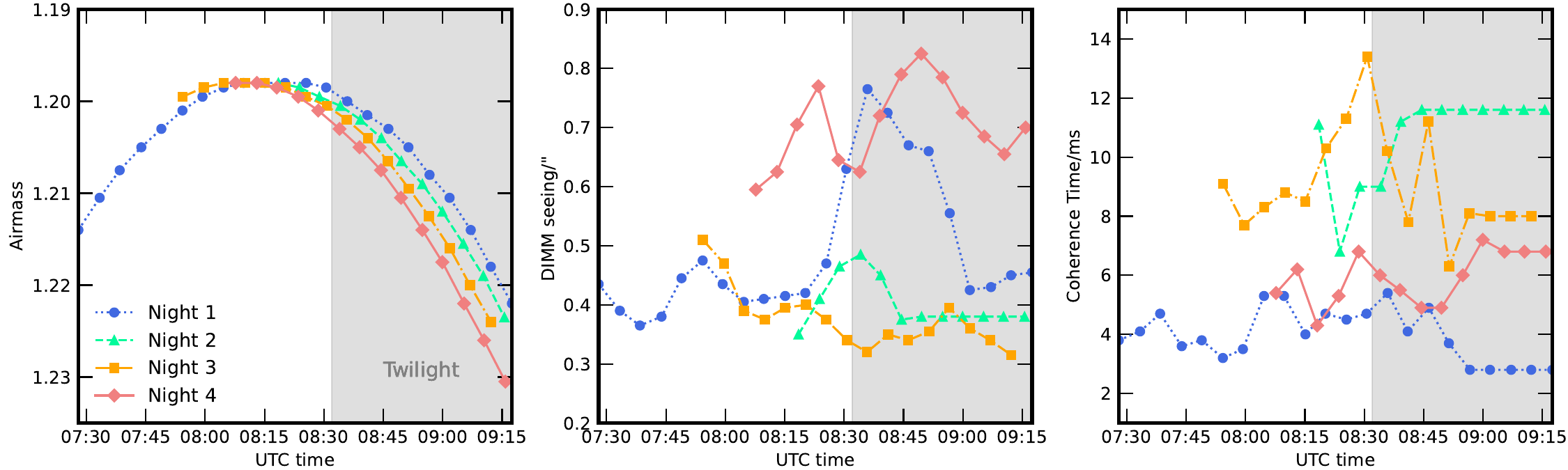}
    \caption{Observing conditions for each night of the observations. The left panel shows the airmass, the middle panel shows the Differential Image Motion Monitor (DIMM) seeing and the right panel shows the coherence time. Note that our observations towards the end of each night were performed in twilight.}
    \label{fig:obs_conditions}
\end{figure*}

We observed AB~Pic~b over four consecutive nights, determining the spectral variability of the data over each of the nights and performing atmospheric retrievals to explore differences in the temperature profile, atmospheric composition and cloud properties on the timescale of $\sim$days. We use the variations in the spectral line shapes and depths over each of the nights to explore this variability, as high-resolution spectroscopy is most reliable when exploring the individual lines and their ratios rather than continuum fluxes \citep[e.g.][]{pelletier2023, maguire2024}. Our observations also allow us to constrain the carbon isotopes given the strong CO absorption features in the K-band. In addition to the chemical and physical properties of the atmosphere, the R$\sim$100,000 observations allow us to determine the rotational velocity and radial velocity of AB~Pic~b, providing an excellent opportunity to probe the orbital solution of the companion.

The next section discusses the methodology, describing the data analysis and atmospheric modelling, followed by the results and discussion for each night of observation individually as well as all four nights combined. Finally, we present our conclusions in section 4.

\section{Methods}
We observed AB~Pic~b with CRIRES+ \citep{kaeufl2004, dorn2014, dorn2023} on the VLT's UT3 as part of the ESO SupJup survey. We performed atmospheric retrievals using the HyDRA retrieval framework \citep{gandhi2019_hydrah, gandhi2023_vhs}. Below, we discuss the observations of AB~Pic~b, the reduction and telluric correction of the data and the modelling setup for the atmospheric retrievals.

\begin{figure}
	\rotatebox[origin=c]{-5}{\includegraphics[width=\columnwidth]{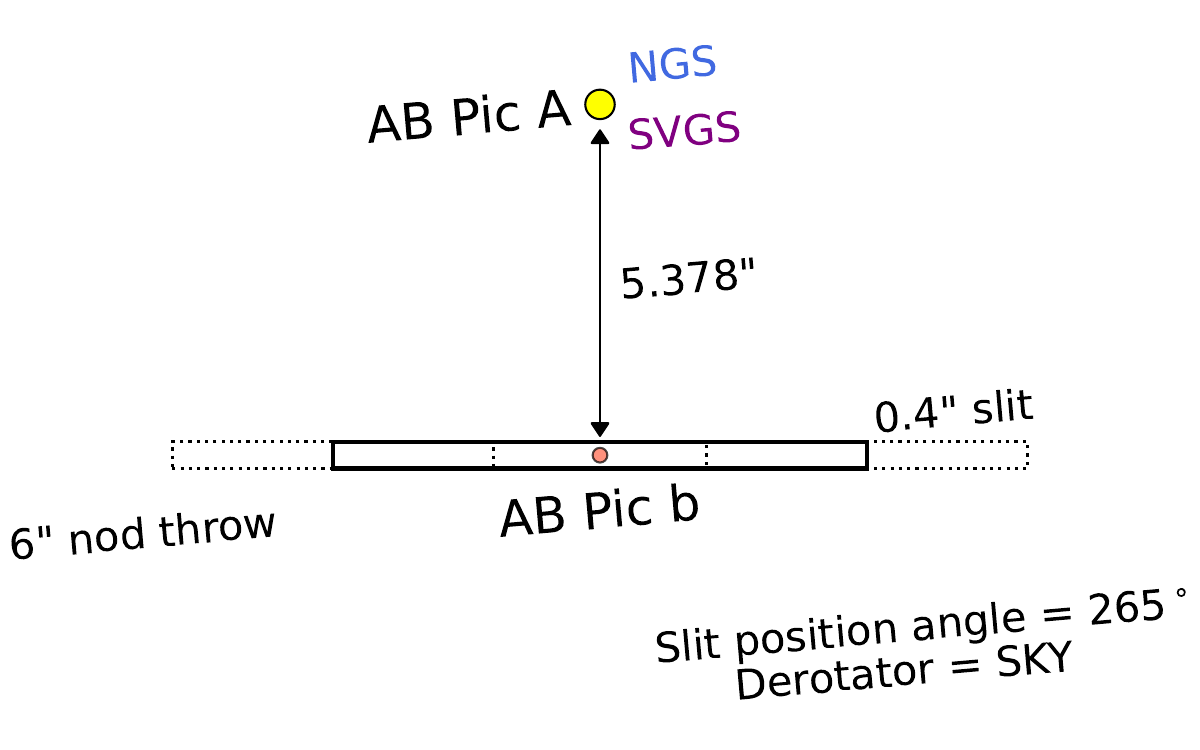}}
    \caption{Set up of the slit orientation for our observations. We used the 0.4" slit, with AB Pic A set as the slit viewer guide star and the natural guide star for the wavefront sensor. We only placed AB~Pic~b on the slit, with a nod throw of 6".}
    \label{fig:slit}
\end{figure}

\begin{table*}
    \centering
    \def\arraystretch{1.5}
\begin{tabular}{c|c|c|c|c|c|c}
\textbf{Night} & \textbf{AB pairs} & \textbf{Observing time/min} & \textbf{DIMM Seeing / \"} & \textbf{Coherence Time / ms} & \textbf{PWV / mm} & \textbf{Spectral resolution} \\
\hline
2022-11-01 & 11 & 110 & 0.365-0.765 & 2.8-5.4 & 1.70-1.74 & 112,800\\
2022-11-02 & 6  & 60 & 0.35-0.485  & 6.8-11.6 & 1.16-1.22 & 97,400\\
2022-11-03 & 8  & 80 & 0.315-0.51  & 6.3-13.4 & 1.26-1.29 & 103,700\\
2022-11-04 & 7  & 70 & 0.595-0.825 & 4.3-7.2  & 2.47-2.79 & 93,200\\
\end{tabular}
    \caption{Observing conditions for the 4 nights. The PWV indicates the precipitable water vapour.}
    \label{tab:nights}
\end{table*}

\subsection{Observations}
AB~Pic~b was observed across four consecutive nights between 01-11-2022 to 04-11-2022, with each of the observations performed towards the end of the night when AB~Pic system was at its lowest airmass (See Figure~\ref{fig:obs_conditions} and Table~\ref{tab:nights}). For all of the nights the seeing was relatively good ($\lesssim 0.8$") and the coherence times were high, particularly for nights two and three. This resulted in a clear signal of the companion, uncontaminated by starlight from AB~Pic~A given the $\sim5.4$" separation, and with a good AO correction (see below). We observed AB~Pic~b into morning twilight for each night as well, but given the K-band observations this did not significantly impact the data quality. Our longest set of observations was for the first night, with 11 AB pairs, and our shortest was for night 2 with 6 AB pairs. We adopted a 300 second exposure (DIT), with NDIT set to 1 for every set of observations.

Figure~\ref{fig:slit} shows the slit viewer position and orientation with respect to the system. We used the 0.4" slit for the observations, with a 6" nod throw, centred on the companion AB~Pic~b. This allowed us to maximise the companion's light entering the slit. We adopted the SKY position for the derotator to ensure the position of the companion and the host star remained the same throughout the observations. We observed AB Pic b alone due to the large on-sky separation between its host star, which made placing both on the slit difficult. This configuration also allowed us to adopt longer exposure times which reduced the relative readout noise. We used the host AB~Pic~A, a 13.3~Myr old K1V star with H-band magnitude 7.09, as both the natural guide star for the adaptive optics and the slit viewer guide star given its brightness. The high coherence times and good seeing resulted in the actual spectral resolution derived for each night (shown in Table~\ref{tab:nights}) being significantly higher than the quoted R$\sim$48,000 for K2166 observations with CRIRES+. We calculated the spectral resolution of our observations by using the width of the trace on the detector, and assumed that the PSF was the same in all directions, as previously done by \citet{holmberg2022} and \citet{nortmann2024}. Each of the nights also had low precipitable water vapour, with relative humidity $\lesssim5\%$, resulting in low telluric H$_2$O contamination.

\subsubsection{Data Reduction}

We reduced the data using the \texttt{excalibuhr}\footnote{\url{https://github.com/yapenzhang/excalibuhr}} pipeline \citep{zhang2024}, a reduction pipeline for CRIRES+ observations of transiting and directly-imaged companions \citep[see also][]{holmberg2022}. The first step of the pipeline is to create the master dark frames, flat, readout noise and bad pixel map from the calibration frames. The traces of the spectral orders are then identified and the slit curvature is corrected using the Fabry-P\'erot Etalon (FPET) frames. The flux of the flat frame is then extracted and used to obtain the blaze function. We then subtract each AB nod pair to remove the sky background and dark current, and the normalised flat field is applied to each of the science frames. Each A and B nod position is then mean combined, and the tiled orders and slits in detector images are straightened. The 1D spectrum is extracted using an optimal extraction algorithm \citep{horne1986} using an extraction aperture of 15 pixels. Note that the B position trace for the shortest wavelength observations is very close to the edge of the detector, but this was not significant for our observations as we only used the reddest 5 orders (see below).

\subsubsection{Telluric Standard}\label{sec:telluric_standard}
\begin{figure*}
	\includegraphics[width=\textwidth]{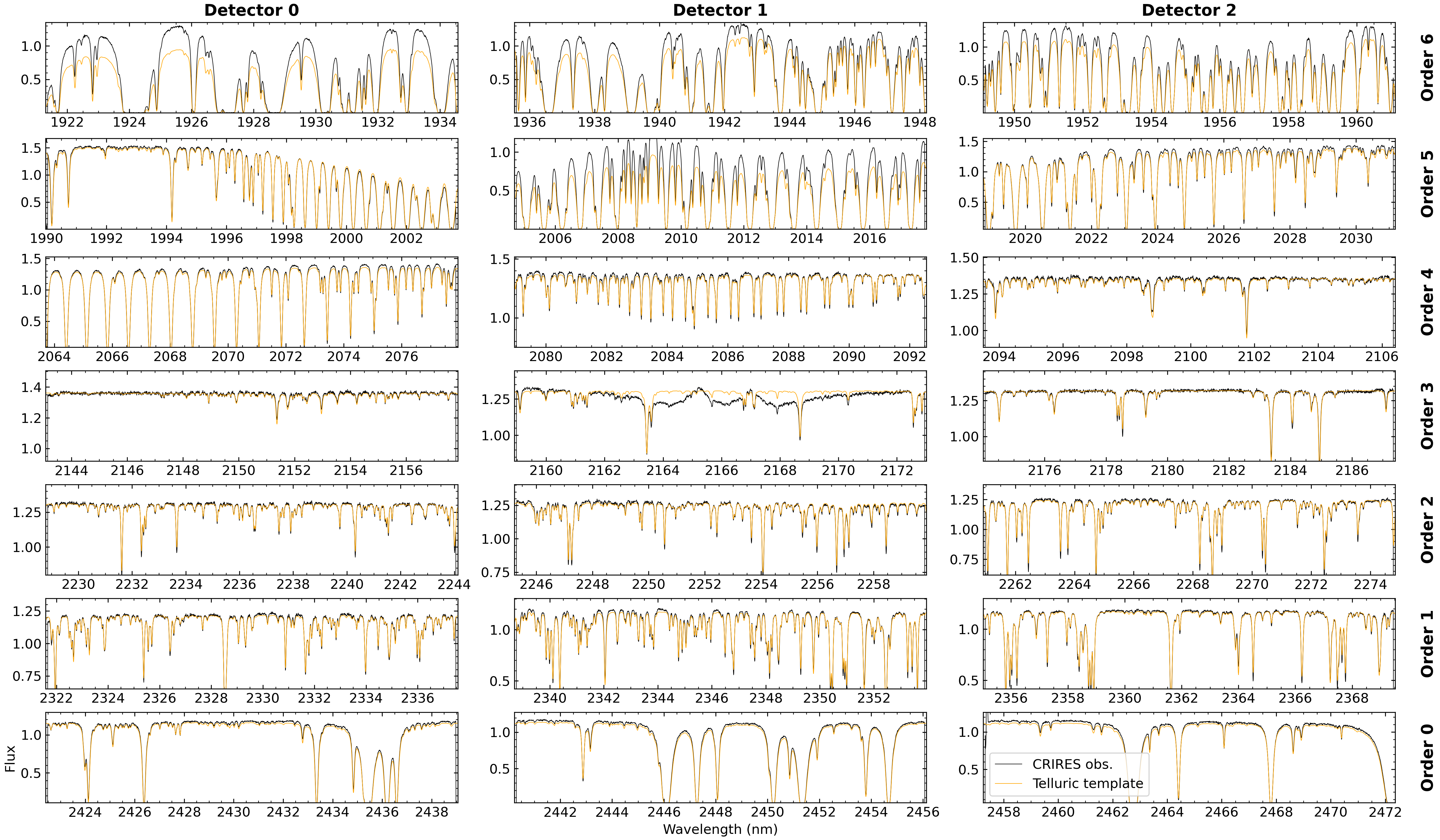}
    \caption{Extracted spectrum of our standard star, $\mu$~Pic, for all seven orders and three detectors for night 1 (2022-11-01). We also show the telluric template used to fit the wavelength solution. Note that detector 1 of order 3 contains the Brackett $\gamma$ line of $\mu$~Pic, which was masked in our analysis.}
    \label{fig:spec_standard}
\end{figure*}

To correct for the telluric absorption we used standard stars, which were $\mu$~Pictoris the first night, 53 Psc for the second, $\lambda$ tau for the third, and $\zeta$ Phe for the final night. These were used to refine the wavelength solution given that the actual observations of AB~Pic~b did not have sufficient signal-to-noise per pixel for high accuracy wavelength calibrations. We use a SkyCalc model \citep{noll2012, jones2013} and applied a third order polynomial on the initial solution from the extraction of the standard star, and used this refined wavelength grid for our AB~Pic~b observations. Figure~\ref{fig:spec_standard} shows the fitted telluric spectrum with the SkyCalc model and the observations of $\mu$~Pic taken on the first night. The spectral features show excellent agreement with the SkyCalc model, with the only difference occurring for the bluest order where there is heavy telluric contamination and the 2161-2171~nm range of the Brackett $\gamma$ line, both of which were masked out for our analysis.

\subsubsection{Extracted Spectrum}

To calibrate the flux we adopted the measured 2MASS and VLT/NACO brightness of K=14.14 \citep{cutri2003, chauvin2005} and GAIA distance of 50.14~pc \citep{gaia2020}. We masked out the regions with heavy telluric contamination to avoid interference in the atmospheric retrievals. We also masked out the Brackett $\gamma$ line in order 3. This masking was carried out independently for each night, as the seeing, precipitable water vapour and average airmass all varied from night to night.

\subsection{Atmospheric Modelling}

We perform atmospheric retrieval analysis on the spectrum of AB~Pic~b to determine the chemical composition and temperature profile of the atmosphere for each of the four nights. This uses the the HyDRA retrieval framework \citep[e.g.][]{gandhi2019_hydrah, gandhi2022}. This has previously been adapted to retrieve the atmospheres of non-irradiated atmospheres \citep{gandhi2023_vhs}.

\subsubsection{P-T Profile}

We use the method of \citet{deregt2024} to determine the P-T profile of the atmosphere, using a grid of P-T points and weighting the 3rd order derivative to reduce ringing/oscillatory behaviour. This is a likelihood penalty imposed for temperature profiles with high 3rd order derivatives \citep[see][]{li2022}. This is given by
\begin{align}
    \ln\mathcal{L}_\mathrm{penalty} &= -\frac{\mathrm{PEN}^{(3)}_\mathrm{gps}}{2\gamma}  - \frac{1}{2}\ln(2\pi\gamma) = -\frac{1}{2\gamma} \|\mathbf{D}_3\mathbf{C}\|^2 - \frac{1}{2}\ln(2\pi\gamma), \label{eq:ln_L_pen}
\end{align}
where $\mathbf{D}_3\mathbf{C}$ is the 3rd order general difference matrix, which computes the P-T points on the knots and weights them by their separation in log pressure for our unevenly spaced P-T points \citep[see][]{gandhi2023_vhs}. We chose the 3rd order derivative to apply the penalty as we expect some curvature of the temperature profile with pressure, as otherwise this would be suppressed/rejected in our Bayesian analysis. The P-T knots we chose are not evenly spaced, and sample the expected photosphere around 1 bar more closely. Similarly to \citet{line2015}, we include the factor $\gamma$ as a free parameter in the retrievals. This parameter divides the penalty applied to the 3rd order derivative, and so allows the likelihood penalty to be suppressed at large $\gamma$ if the data warrants a temperature profile with more oscillatory features. Hence we have 9 free parameters for the temperature profile, with the temperature retrieved at various pressures throughout the atmosphere (see Table~\ref{tab:priors}), and an additional parameter for the P-T penalty factor $\gamma$.

\subsubsection{Atmospheric Opacity}
The atmospheric model includes opacity arising from a range of sources. Firstly, we include the line absorption from the molecular species in the atmosphere. This uses a line list to calculate the cross section for each species. We use the ExoMol line list database \citep{tennyson2016} for H$_2$O \citep{polyansky2018}, HCN \citep{harris2006, barber2014} and NH$_3$ \citep{coles2019}, and the HITEMP database \citep{rothman2010} for CH$_4$ \citep{hargreaves2020} and CO and its isotopologues \citep{li2015}. We also use the line list from \citet{huang2013} and \citet{huang2017} for CO$_2$. For each pressure and temperature grid point every spectral line is broadened into a Voigt profile \citep[e.g.][]{gandhi2017} using the H$_2$ and He pressure broadening coefficients \citep{gandhi2020_cs}. This is then summed for every line to give the overall cross section from each species. We retrieve the volume mixing ratio of each chemical species as a free parameter, with vertically constant abundance profiles. This has been shown to be the most optimal setup from previous work \citep{deregt2024}. In addition to molecular line absorption, we also include opacity arising from collisionally induced absorption due to H$_2$-H$_2$ and H$_2$-He interactions \citep{richard2012}.

In addition to the opacity due to gaseous chemical species, we also include a cloud opacity model based on \citet{molliere2020}. The cloud opacity $\kappa_\mathrm{cl}$ at a pressure $P$ and wavelength $\lambda$ is given by 
\begin{align}
        \kappa_\mathrm{cl}(P, \lambda) = 
    \begin{cases}
      \kappa_{\mathrm{cl},0}(\lambda) \left(\dfrac{P}{P_\mathrm{cl}}\right)^{\alpha_\mathrm{cl}} & P\leq P_\mathrm{cl}, \\
      0 & P> P_\mathrm{cl}, \\
    \end{cases}
\end{align}
with $P_\mathrm{cl}$ and $\alpha_\mathrm{cl}$ representing the cloud deck pressure and power law respectively, both of which are set to free parameters for the retrieval. We determine $\kappa_{\mathrm{cl},0}(\lambda)$ through
\begin{align}
    \log(\kappa_{\mathrm{cl},0}(\lambda)) = \log(&\kappa_{\mathrm{cl, 2\mu m}}) \, +\nonumber \\
    & \frac{\lambda - \mathrm{2\mu m}}{\mathrm{2.5 \mu m} - \mathrm{2 \mu m}} (\log(\kappa_{\mathrm{cl, 2.5\mu m}}) - \log(\kappa_{\mathrm{cl, 2\mu m}})),
\end{align}
where $\kappa_{\mathrm{cl, 2\mu m}}$ and $\kappa_{\mathrm{cl, 2.5\mu m}}$ are free parameters in the retrieval. The opacity at intermediate wavelengths is then interpolated between 2~$\mu$m and 2.5~$\mu$m.

\subsubsection{Likelihood Evaluation}

We adopt a method similar to \citet{ruffio2019} and \citet{gibson2020} to determine the likelihood \citep[see][for further details]{deregt2024}. For each order and detector, we construct the covariance matrix $\mathbf{\Sigma}$ as
\begin{align}
    \Sigma_{ij} = \delta_{ij}\sigma_{i}^2 + a^2 \sigma_{\mathrm{eff},ij}^2 \exp\Big(-\frac{r_{ij}^2}{2 l^2}\Big),
\end{align}
with $a$ and $l$ representing the gaussian process amplitude and lengthscale, which are both free parameters in the retrieval. The diagonal elements consist of the overall error $\sigma_i$ for each data point, and the off diagonal elements give the correlated component. Here,
\begin{align}
    r_{ij} = |\lambda_i - \lambda_j|.
\end{align}
We also define the arithmetic mean of the variances of elements $i$ and $j$,
\begin{align}
    \sigma_{\mathrm{eff},ij}^2 = \frac{1}{2} (\sigma_i^2 + \sigma_j^2).
\end{align}

The overall log likelihood for a covariance matrix $\mathbf{\Sigma}$ is then given by
\begin{align}
    \ln \mathcal{L} = -\frac{1}{2} N \ln(2\pi) -& \frac{1}{2} \ln(\mathrm{det}(\mathbf{\Sigma}))\nonumber \\
    & - \frac{1}{2} N \ln(s^2)- \frac{1}{2 s^2} \mathbf{r}^T \mathbf{\Sigma}^{-1} \mathbf{r},
\end{align}
where $N$ is the number of data pofints. The residual $\mathbf{r}$ is defined
\begin{align}
    \mathbf{r} = \mathbf{d} - \mathbf{m}\phi,
\end{align}
where $\mathbf{d}$ refers to the data, $\mathbf{m}$ refers to the model, and the scale factor $\phi$ that is applied to the model for each detector and order except for the first \citep[see][]{deregt2024}. This scale factor is calculated through
\begin{align}
    \phi = \frac{\mathbf{m}^T \mathbf{\Sigma}^{-1} \mathbf{d}}{\mathbf{m}^T \mathbf{\Sigma}^{-1} \mathbf{m}}.
\end{align}
The $s^2$ term represents a scaling of the overall covariance for each order and detector. This is given by
\begin{align}
    s^2 = \frac{1}{N}\mathbf{r}^T \mathbf{\Sigma}^{-1} \mathbf{r}.
\end{align}

\begin{table}
    \centering
    \begin{tabular}{c|c|c}
& \textbf{Parameter}              & \textbf{Prior Range}\\
\hline
Chemistry   & $\log(\mathrm{H_2O})$  & -12 $\rightarrow$ -1 \\
            & $\log(\mathrm{^{12}CO})$ & -12 $\rightarrow$ -1 \\
            & $\log(\mathrm{^{13}CO})$ & -12 $\rightarrow$ -1 \\
            & $\log(\mathrm{C^{18}O})$ & -12 $\rightarrow$ -1 \\
            & $\log(\mathrm{HCN})$ & -12 $\rightarrow$ -1 \\
            & $\log(\mathrm{CH_4})$ & -12 $\rightarrow$ -1 \\
            & $\log(\mathrm{NH_3})$ & -12 $\rightarrow$ -1 \\
            & $\log(\mathrm{CO_2})$ & -12 $\rightarrow$ -1 \\
\hline
Temp. Profile & $T_\mathrm{100bar}$ / K & 300 $\rightarrow$ 4000 \\
            & $T_\mathrm{10bar}$ / K & 300 $\rightarrow$ 4000 \\
            & $T_\mathrm{3bar}$ / K & 300 $\rightarrow$ 4000 \\
            & $T_\mathrm{1bar}$ / K & 300 $\rightarrow$ 4000 \\
            & $T_\mathrm{0.3bar}$ / K & 300 $\rightarrow$ 4000 \\
            & $T_\mathrm{0.1bar}$ / K & 300 $\rightarrow$ 4000 \\
            & $T_\mathrm{0.01bar}$ / K & 300 $\rightarrow$ 4000 \\
            & $T_\mathrm{10^{-4}bar}$ / K & 300 $\rightarrow$ 4000 \\
            & $T_\mathrm{10^{-6}bar}$ / K & 300 $\rightarrow$ 4000 \\
\hline
P-T penalty & $\log(\gamma)$ & -3 $\rightarrow$ 2 \\
\hline
Clouds/hazes & $\log(\kappa_\mathrm{cl, 2\mu m}/\mathrm{cm^2g^{-1}})$ & -10 $\rightarrow$ 50 \\ 
            & $\log(\kappa_\mathrm{cl, 2.5\mu m}/\mathrm{cm^2g^{-1}})$ & -10 $\rightarrow$ 50 \\ 
            & $\log(P_\mathrm{cl}/\mathrm{bar})$ & -4 $\rightarrow$ 2 \\
            & $\log(\alpha_\mathrm{cl})$ & 0 $\rightarrow$ 20 \\
\hline 
GP Parameters & $^\dagger$$\log(a)$ & -2.0 $\rightarrow$ 0.3 \\
            & $\log(l/\mathrm{\mu m})$ & -3.0 $\rightarrow$ -1.4 \\
\hline
Companion params. & $\mathrm{R_p/R_J}$ & 0.6 $\rightarrow$ 2.5 \\
            & $\log(\mathrm{g/cms^{-2}})$ & 3.5 $\rightarrow$ 5 \\
\hline
Rotation    & $v\sin i/\mathrm{km s^{-1}}$ & 1 $\rightarrow$ 50 \\
            & $\mathrm{\epsilon}$ & 0 $\rightarrow$ 1 \\
\hline 
Radial Velocity    & $^\ast$$\mathrm{RV}$ / kms$^{-1}$ & -30 $\rightarrow$ 30 \\

    \end{tabular}
    \caption{Parameters and uniform prior ranges for our retrieval of AB~Pic~b. \\
    $^\dagger$ Each order for each night has its own separate GP amplitude $a$. This results in 5 parameters for $a$ for the 5 orders for each night, and 20 free parameters for $a$ for the combined retrieval across the 4 nights. \\
    $^\ast$ We retrieve separate $\mathrm{RV}$ values for each night in the combined retrieval.}
    \label{tab:priors}
\end{table}

\subsubsection{Retrieval Setup}

We run our retrieval for each night separately, and perform a combined retrieval which analyses the four nights simultaneously. In addition to the chemistry, temperature profile, clouds and GP evaluation, we also retrieve the radius of the companion and the surface gravity, $\log(\mathrm{g})$, as free parameters (see Table~\ref{tab:priors}). This is because for our directly-imaged companion the mass and radius uncertainties are quite large as they are based on evolutionary models \citep{chauvin2005}. We generate our spectral models at a spectral resolution of $R=300,000$, and convolve them down to the instrumental resolution for each night, which ranges between $R=93,200$ and $R=112,800$ (see Table~\ref{tab:nights}). Once the instrumental broadening has been applied to the model, any additional broadening of the spectrum will be due to the companion's rotation. We therefore leave the rotational velocity of AB~Pic~b and the $\epsilon$ (the limb-darkening coefficient) as free parameters. It is therefore important in this case that the actual spectral resolution of the observations be calculated effectively, as otherwise the rotational broadening will be biased. In addition, we retrieve the radial velocity shift of the spectrum due to Doppler motion of Ab~Pic~b. Given slight differences in the barycentric velocity as well as wavelength calibration between nights, we retrieve separate velocity shifts for each night in the combined retrieval of the four nights. We apply the barycentric correction for each night on the spectrum to obtain the $\mathrm{RV}$, and hence this value represents the radial velocity of the companion in the heliocentric frame.

Altogether, we have 33 free parameters for the retrieval on each individual night, and 51 free parameters for the combined retrieval of the four nights together. To obtain the detection significances of each species we also perform separate retrievals with the species removed. For our analysis we only use the reddest five orders, with wavelengths between $\sim2064$~nm and $\sim2475$~nm, given the heavy telluric contamination in the bluest two orders. We perform our retrievals using the MultiNest Nested Sampling algorithm \citep{feroz2008, feroz2009, feroz2013, buchner2014}, with 500 live points.

\subsubsection{High-Pass Filtering}\label{sec:highpass}
During our retrievals we high-pass filter the data and the model to remove any long range wavelength dependence over the order/detector. This ensures that the retrieved constraints are driven entirely by the individual line features of spectrally active species. We adopted this approach because high-resolution spectroscopy is inherently more reliable in obtaining line ratios and features than the continuum \citep[e.g.][]{gandhi2023, maguire2024}, particularly so given that the observing conditions varied over our 4 nights of observations (see Table~\ref{tab:nights}). Hence, it is more difficult to determine whether any deviations in the non high-pass filtered spectra between nights are definitively due to variability and not due to changing observing conditions. We therefore apply a boxcar rolling average which is 100 pixels width to remove the long range structure within each order and detector.

The use of the high-pass filter does affect the constraints on a number of atmospheric parameters from our retrieval. Firstly, the radius of the planet is less certain because we cannot use the continuum flux level to determine its value. The radius is therefore constrained indirectly through line shape variations and the log(g) value, which is also coupled to the chemical abundances of each species. As the continuum level is lost, the absolute photospheric temperatures also become more uncertain. The temperature gradients are intrinsically coupled to chemical abundances given that line features can be fit with higher abundances with shallower temperature gradients or lower abundances with more steep gradients. Therefore, the absolute values of the radius, log(g), temperature and abundances become less precise but remain accurate and reliable. However, abundance ratios are very robust given that the line ratios of the many molecular lines in the observations are not altered, and any changes to the temperature, log(g) or radius affect the lines from all of the species in a similar way. Therefore, compositional ratios such as C/O and $^{12}$C/$^{13}$C will still be tightly constrained even though absolute abundances may not be, as is commonly seen in both ground-based high-resolution spectroscopy and JWST observations with NIRSpec \citep[e.g.][]{pelletier2023, gandhi2023_vhs}.

\begin{figure*}
	\includegraphics[width=\textwidth]{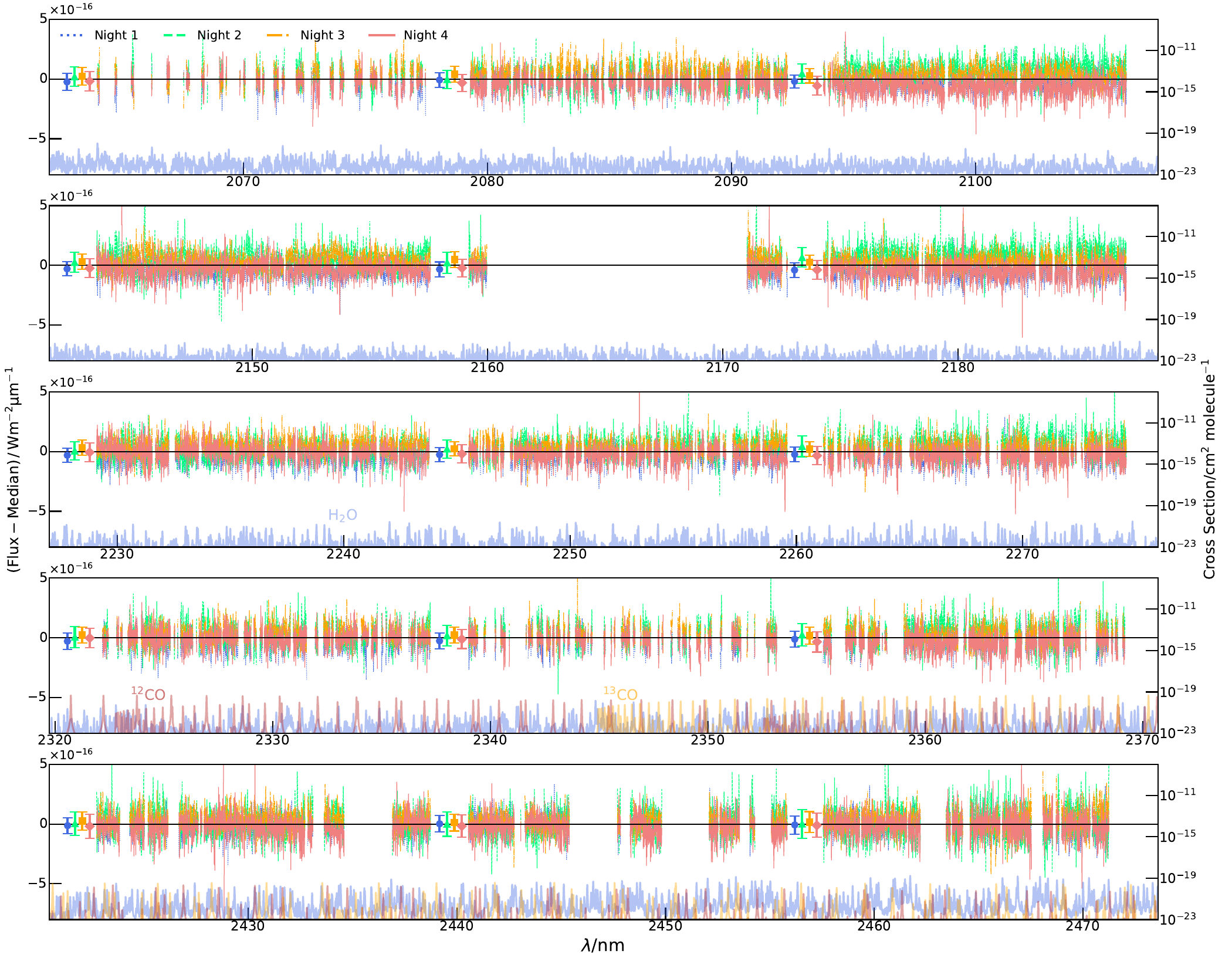}
    \caption{Data for each night subtracted by the median spectrum of all 4 nights. The mean residual for each order and detector is indicated with the associated error bar on the left hand side of each order-detector pair. We also show the molecular cross sections of H$_2$O, $^{12}$CO and $^{13}$CO at the bottom of each of the panels to indicate where features are expected if there are significant variations between nights.}
    \label{fig:median_subtracted_spec}
\end{figure*}

\begin{figure}
	\includegraphics[width=\columnwidth]{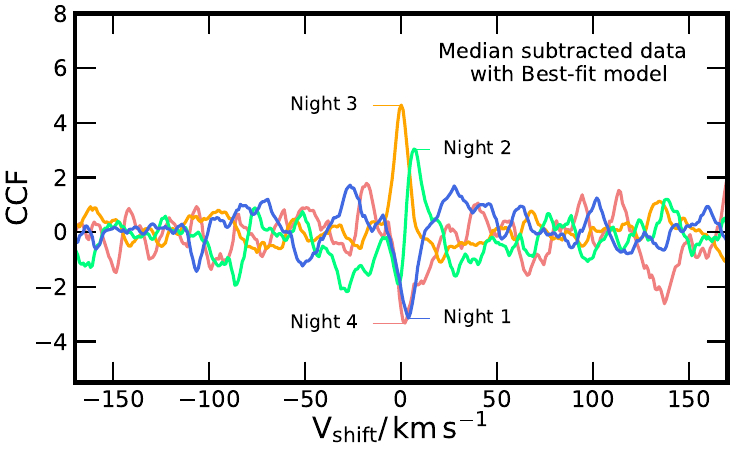}
    \caption{Cross-correlation function (CCF) of the high-pass filtered data for each night subtracted by the median over all nights against the best fit model from the combined retrieval. We divide the the values by the standard deviation away from the peak and hence they act as a proxy for the signal-to-noise. Positive correlation indicates the night had higher line depths/stronger features over the others.}
    \label{fig:ccf_med_subtract}
\end{figure}

\begin{figure*}
	\includegraphics[width=\textwidth]{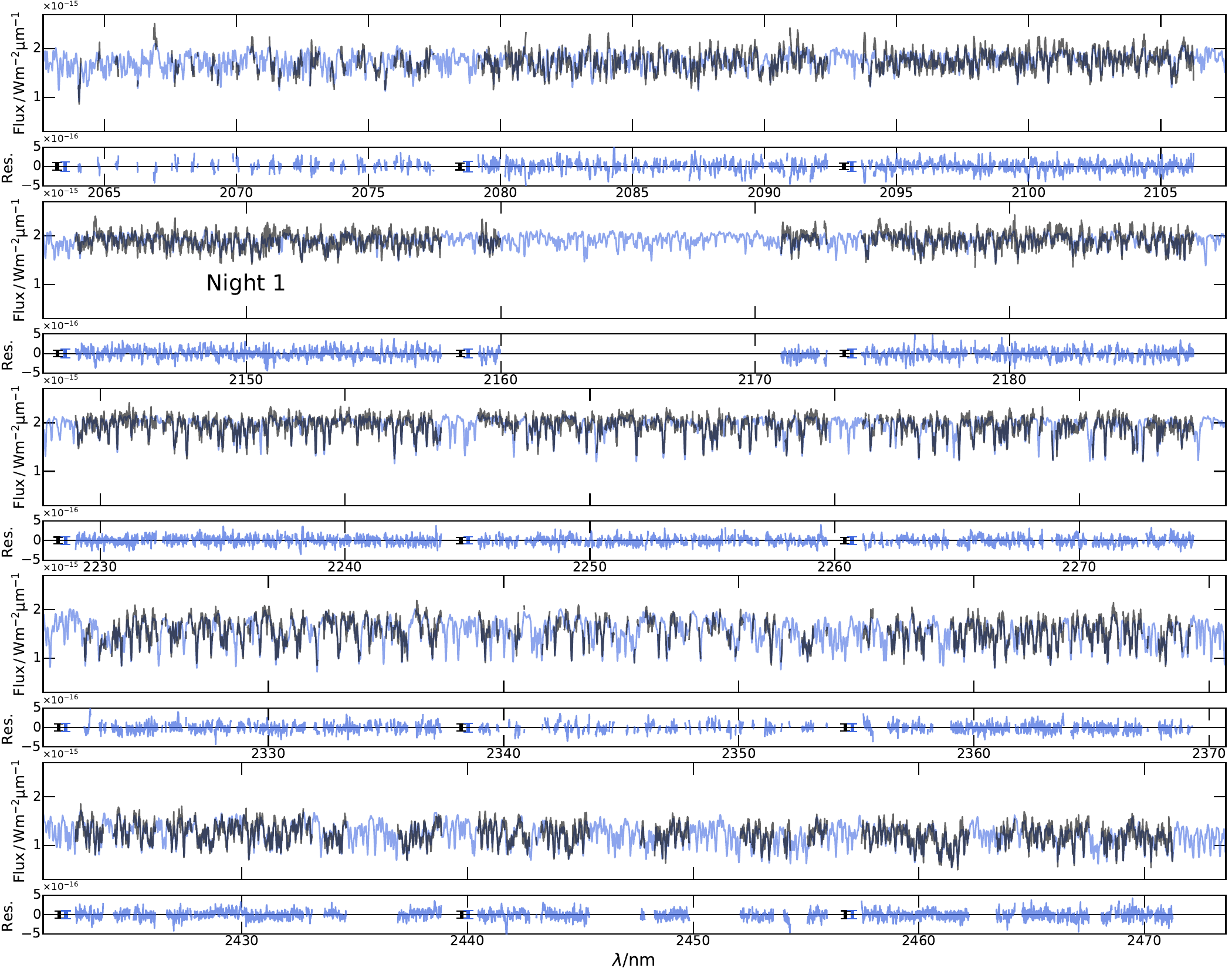}
    \caption{Best fit model from the retrieval of night 1 (blue) against the observations (black). We also show the residuals between the model and the data in the bottom panels for each order. The mean photon noise for each order and detector is indicated with a black error bar, and the standard deviation of the residuals is shown in the blue error bar.}
    \label{fig:best_fit}
\end{figure*}

\begin{figure*}
	\includegraphics[width=\textwidth]{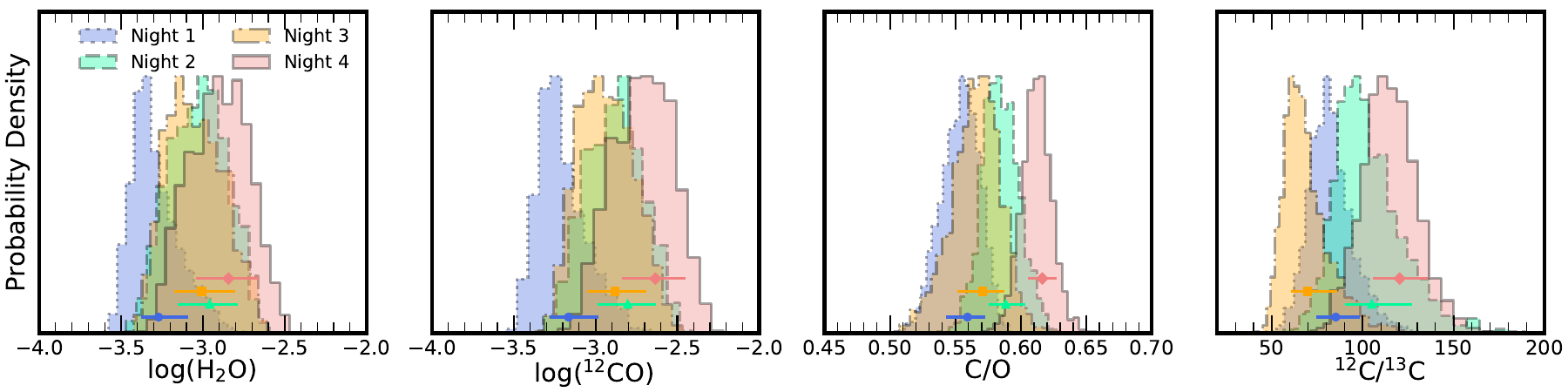}
    \caption{Marginalised posterior distributions of the volume mixing ratios for H$_2$O and CO, and the derived C/O and $^{12}$C/$^{13}$C ratio of the atmosphere of AB~Pic~b for each of the nights. The horizontal error bars indicate the median and $\pm1\sigma$ uncertainties on each constraint.}
    \label{fig:posterior}
\end{figure*}

\begin{table*}
    \centering
    \def\arraystretch{1.5}
\begin{tabular}{c|c|c|c|c|c}
\textbf{Night} & $\mathbf{log(H_2O)}$ & $\mathbf{log(CO)}$ & \textbf{C/O} & \textbf{$\mathbf{^{12}CO/^{13}CO}$} & \textbf{$^{13}$CO det. signif.}\\
\hline
2022-11-01 & $-3.27^{+0.18}_{-0.11}$ & $-3.17^{+0.18}_{-0.11}$ & $0.56^{+0.01}_{-0.02}$ & $85^{+13}_{-11}$ & 6.1$\sigma$\\
2022-11-02 & $-2.96^{+0.17}_{-0.19}$ & $-2.80^{+0.17}_{-0.19}$ & $0.59^{+0.01}_{-0.01}$ & $105^{+22}_{-15}$ & 4.8$\sigma$\\
2022-11-03 & $-3.01^{+0.20}_{-0.17}$ & $-2.88^{+0.19}_{-0.17}$ & $0.57^{+0.02}_{-0.02}$ & $70^{+16}_{-9}$ & 7.2$\sigma$\\
2022-11-04 & $-2.84^{+0.19}_{-0.20}$ & $-2.63^{+0.18}_{-0.21}$ & $0.62^{+0.01}_{-0.01}$ & $120^{+16}_{-15}$ & 6.3$\sigma$\\
\hline
\textbf{Combined Nights} & $-3.15^{+0.08}_{-0.08}$ & $-3.00^{+0.08}_{-0.08}$ & $0.59^{+0.01}_{-0.01}$ & $102^{+8}_{-8}$ & 12.0$\sigma$\\
\end{tabular}
    \caption{Retrieved constraints on the atmospheric composition and the detection significance of $^{13}$CO for each of the 4 nights individually as well as the combined nights.}
    \label{tab:posterior}
\end{table*}

\section{Results and Discussion}

In this section we discuss the constraints on the atmospheric chemistry, temperature profile, cloud deck and other parameters from the retrievals of AB~Pic~b. We first discuss the variation in the spectra over the nights, then report on the retrievals for each night of observations individually. We then discuss the retrieval over all of the nights combined. Finally, we explore the spin axis and obliquity of AB~Pic~b from our retrieved projected rotation velocity and the revisited orbital solution.

\subsection{Spectral line variations in the data between nights}
Figure~\ref{fig:median_subtracted_spec} shows the spectra of each night with the median spectrum combining all 4 nights subtracted. This shows that there are some differences between nights, particularly for the reddest detectors for night 2. This night had the lowest number of AB pairs and hence the lowest signal-to-noise, which may make the blaze correction more likely to be biased. This night also had the lowest spectral resolution, so it is not surprising that this is the night with the largest deviation from the median. The fact that the most strongly affected regions are in the reddest detector for all of the orders indicates that this effect is instrumental, and does not represent night-to-night atmospheric variations. Given that absolute fluxes and continuum values are challenging with ground-based observations at high resolution, it is more difficult to construct a variability in absolute flux over time such as that for other targets with JWST \citep[e.g.][]{biller2024, mccarthy2024}. Therefore, we adopt an approach to look for line depth and shape variations across the nights, which are significantly more stable (see section~\ref{sec:highpass}). This may limit the degree of variability we are sensitive to, but it should provide the most accurate constraints even though their uncertainty may be higher.

The variations in the line features themselves are small and therefore more difficult to detect. We looked for differences in the spectral lines between each night by subtracting the median from the high-pass filtered data, and performing a cross-correlation against our best-fit spectral model, shown in Figure~\ref{fig:ccf_med_subtract}. This ensures that we are sensitive to the variations of the individual spectral lines, and the use of this cross-correlation function (CCF) offers a powerful tool to simultaneously explore many thousands of spectral lines at once. For night 3 we see a positive correlation, indicating that the spectral line features are deeper for this night, and vice versa for nights 1 and 4. Therefore, the data do point to variations in the spectra between nights, albeit with a relatively modest $|\mathrm{CCF}|\lesssim4$ for each night. Note that if some lines are stronger but others weaker then the overall cross-correlation will be average over these, which may underestimate the variability. On the other hand, using the median-subtracted high-pass filtered data is also prone to errors if there are strong differences in the spectral resolution and seeing between the nights, which will overestimate the variations in the spectra. To obtain a more complete picture of the atmosphere of AB~Pic~b we therefore retrieve the various atmospheric parameters for each night.

\subsection{Atmospheric Retrieval of the Individual Nights}

Figure~\ref{fig:best_fit} shows the best fitting atmospheric retrieval model against the data for night 1, with the other nights shown in the appendix section~\ref{sec:best_fit_appendix}. These show an excellent overall fit to all of the 5 orders we used for our analysis, and therefore that our model is flexible enough to capture the spectral features well. The residuals also indicate that there are no significant lines which are not well fit and hence that the sources of opacity which we have included in our model are sufficient. Note that we have masked regions with heavy telluric contamination, in particular the second order with the broad Brackett $\gamma$ line from the telluric standard, as discussed in section~\ref{sec:telluric_standard}. 

Figure~\ref{fig:posterior} and Table~\ref{tab:posterior} show the constraints on the volume mixing ratios for H$_2$O and $^{12}$CO, and the derived values for the C/O, and the $^{12}$C/$^{13}$C ratio from the $^{13}$CO abundance constraints. These show good consistency across the nights, but there is some variation on the order of $\sim$1-2$\sigma$ for the chemical abundances, and $\sim3\sigma$ for the C/O and isotope ratios. Night 1 shows the lowest abundances for both H$_2$O and $^{12}$CO, and night 4 shows the highest. All appear to have broadly similar uncertainties, but night 1 has the highest precision given its longest observing time with 11 AB pairs (see Table~\ref{tab:nights}). While the absolute values of the abundances have a wide uncertainty, the C/O and isotope ratios are relatively tightly constrained given that we are more sensitive to the relative line depth between species rather than the absolute value given the high-pass filter applied to the data. The C/O ratio is largely determined by the CO and H$_2$O, as these are the dominant species of the atmosphere. This is consistent with chemical models at such temperatures \citep[e.g.][]{lodders2002, madhu2012}, which show that CO is the dominant carrier of carbon and CO and H$_2$O are the dominant carriers of oxygen at temperatures $\gtrsim$1200~K. We included HCN, CH$_4$, NH$_3$ and CO$_2$ in our models but these did not show any significant constraints for any of the nights.

\begin{figure}
	\includegraphics[width=\columnwidth]{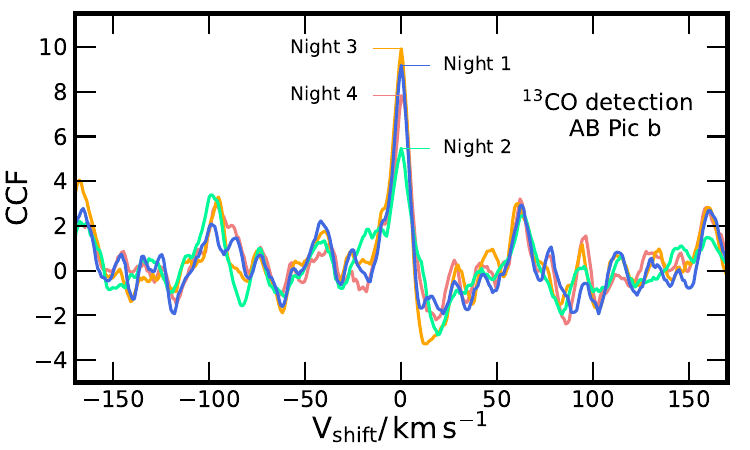}
    \caption{Cross-correlation function (CCF) for $^{13}$CO for each of the individual nights of observations for AB~Pic~b. We divide the the values by the standard deviation away from the peak and hence they act as a proxy for the signal-to-noise. We performed the cross-correlation on the observed spectrum centred in its rest frame.}
    \label{fig:ccf}
\end{figure}

\begin{figure}
	\includegraphics[width=\columnwidth]{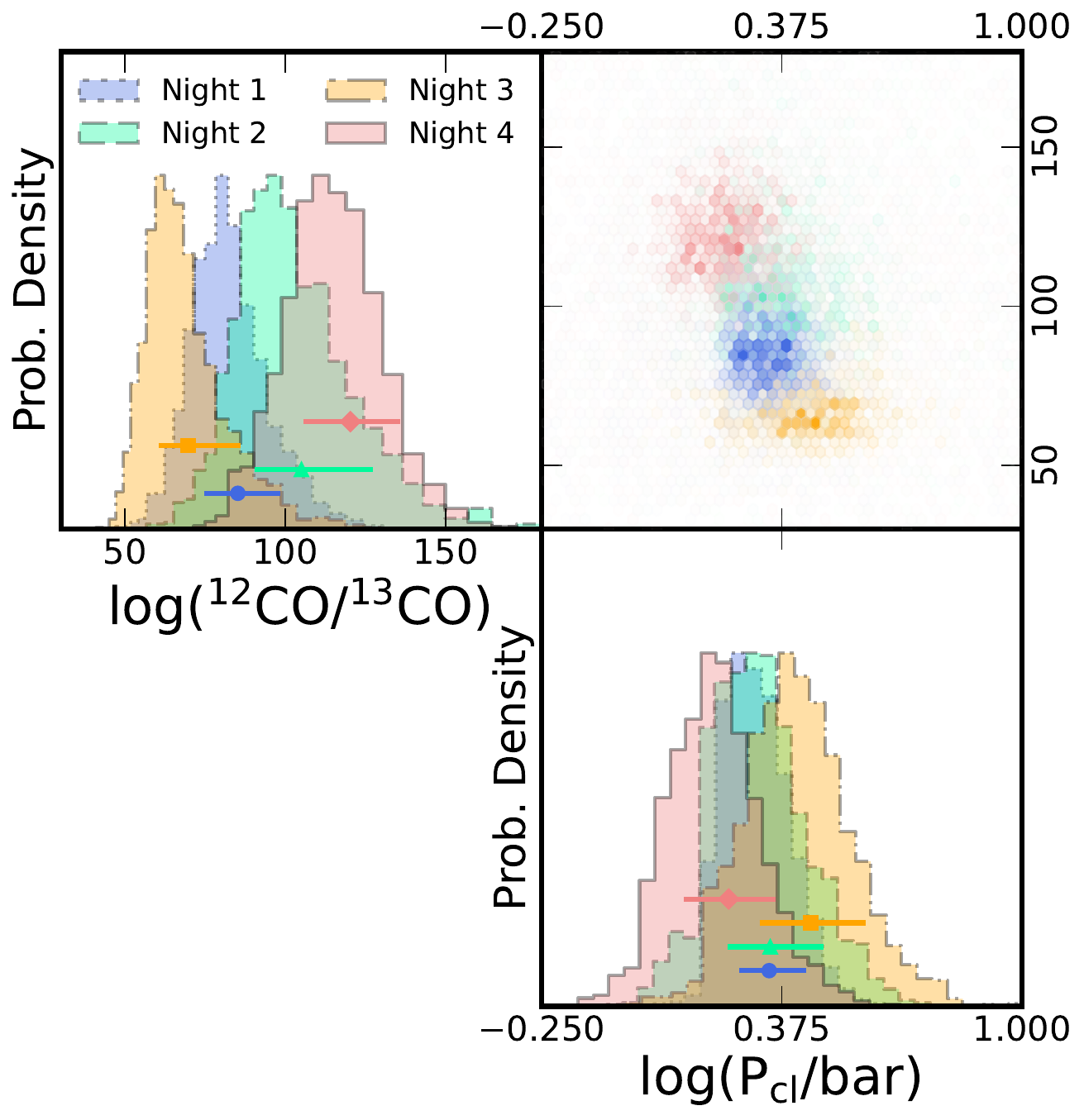}
    \caption{Posterior distribution for the $^{12}$CO/$^{13}$CO abundances and cloud deck pressure for each night observed, showing the correlation between these two parameters.}
    \label{fig:12_13CO_cloud}
\end{figure}

\subsubsection{$^{12}$C/$^{13}$C ratio}\label{sec:isotopes}

The $^{13}$CO isotopes are strongly detected at $>4.5\sigma$ for all nights, as shown in Table~\ref{tab:posterior}. We verified our detection of $^{13}$CO through a cross-correlation function \citep[see e.g.][]{zhang2021_bd, deregt2024}, as shown in Figure~\ref{fig:ccf}. This confirms that the detection of $^{13}$CO for each night is robust, and follows a similar trend as the detections from the retrievals, with night 3 having the strongest detection and night 2 having the weakest. This is despite the first night, night 1, having a much longer observing time. Such differences in the detection significance and isotope ratio could be driven by patchy clouds in the atmosphere, which may be varying on the timescale of $\sim$1 day \citep[e.g.][]{tan2021, lee2024} and affecting the detectability of $^{13}$CO. This also backs up the cross-correlation of the median subtracted data in Figure~\ref{fig:ccf_med_subtract}, which showed that night 3 has slightly stronger spectral lines, potentially due to a partially clearer atmosphere with a deeper cloud deck, thereby increasing the line strengths of trace species such as $^{13}$CO.

\begin{figure}
	\includegraphics[width=\columnwidth]{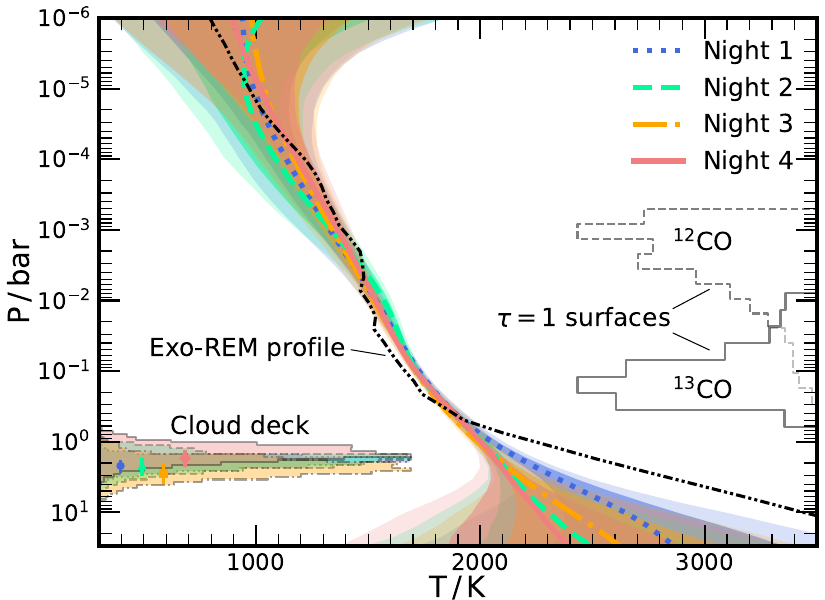}
    \caption{Retrieved Pressure-Temperature (P-T) profile for each of the 4 nights of observations. We show the median best-fit profile with the lines, and the dark and light shaded regions show the 1$\sigma$ and 2$\sigma$ uncertainty ranges respectively. We also show the P-T profile from the best-fit solution derived in \citet{palma-bifani2023} from Exo-REM self-consistent models \citep{charnay2018}. On the left hand side we also plot the constraints on the cloud deck for each night, and on the right hand side we show the locations of the $\tau=1$ surface histograms for $^{12}$CO and $^{13}$CO, indicating the pressures which each species is most sensitive to.}
    \label{fig:PT}
\end{figure}

We do see differences in the constrained isotopologue constraints over our four nights of observations (see Figure~\ref{fig:posterior}). Night 3 shows the lowest $^{12}$CO/$^{13}$CO ratio of all of the nights, $^{12}$CO/$^{13}$CO~$ = 70^{+16}_{-9}$ and night 4 shows the highest, with $^{12}$CO/$^{13}$CO~$ = 120^{+16}_{-15}$. Figure~\ref{fig:12_13CO_cloud} also seems to indicate a correlation between the $^{12}$CO/$^{13}$CO abundance and the cloud deck pressure across the four nights. Night 3 constrains the cloud deck at a higher pressure than the other nights, while simultaneously constraining the lowest $^{12}$CO/$^{13}$CO abundance. If on night 3 the atmosphere was less cloudy and we were more sensitive to higher pressures where the $^{13}$CO lines are generated, the constrained $^{13}$CO abundance may be higher and hence reduce the $^{12}$C/$^{13}$C ratio. At the same time, an increased sensitivity to the higher pressures would also increase the detection significance as we see for this night, and explain the line depths being strongest for this night in Figure~\ref{fig:ccf_med_subtract}. The $^{13}$CO would be more susceptible to the effect of clouds given these lines are generated at deeper pressures than the $^{12}$CO. The $\tau=1$ surface histograms in Figure~\ref{fig:PT} confirm that $^{13}$CO is most sensitive to pressures only $\lesssim1$~dex above the cloud deck, whereas the $^{12}$CO is most sensitive to $\sim$mbar pressures. However, we should note that different spectral noise realisations and/or the effectiveness of the telluric fit for each of the nights may also contribute to the observed difference to the isotopologue ratios between the nights. Therefore, more observations and modelling are required to definitively assess if this trend is significant and if patchy/broken clouds are present in the atmosphere.

\begin{figure}
	\includegraphics[width=\columnwidth]{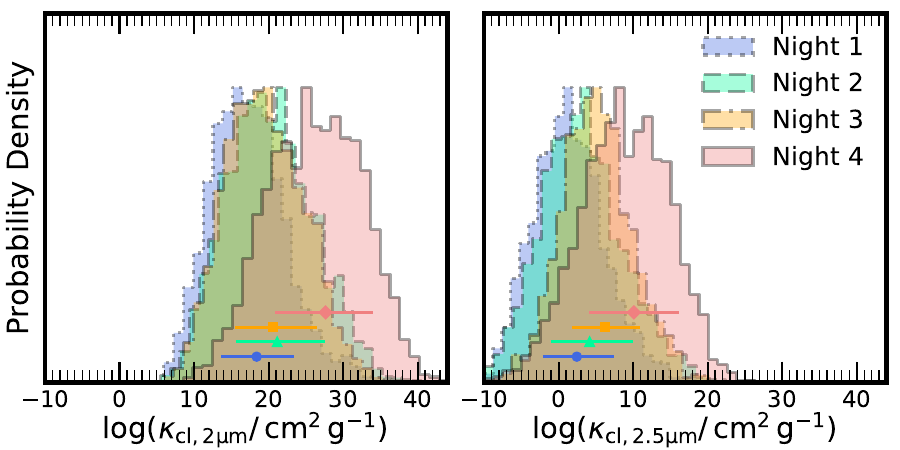}
    \caption{Marginalised posterior distributions for the cloud opacity at 2~$\mu$m and 2.5~$\mu$m for the individual nights of observation for AB~Pic~b.}
    \label{fig:posterior_clouds}
\end{figure}

\begin{figure*}
	\includegraphics[width=\textwidth]{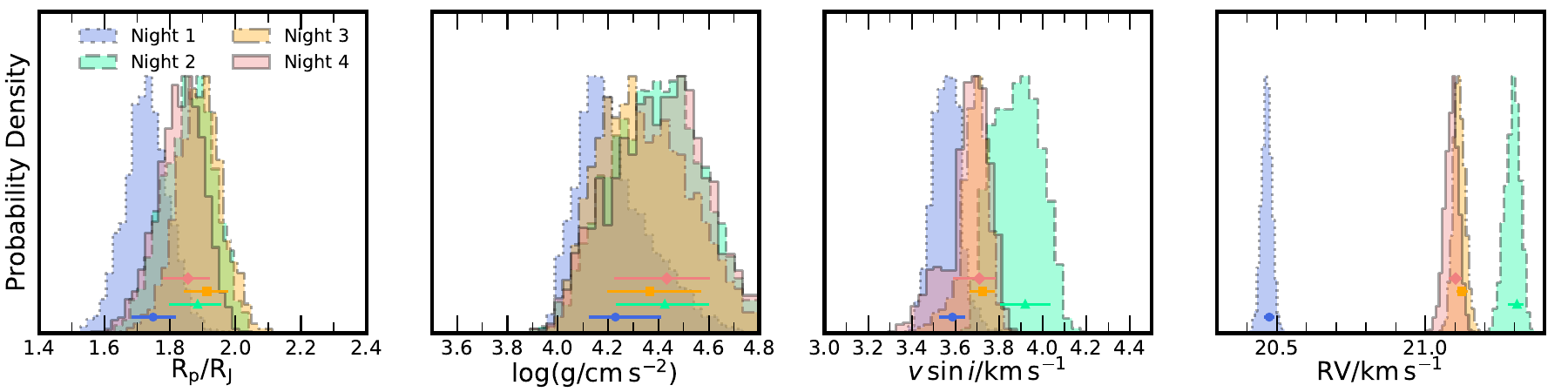}
    \caption{Marginalised posterior distributions for the radius, log(g), $v\sin i$ and $\mathrm{RV}$ for the individual nights of observation for AB~Pic~b.}
    \label{fig:posterior_other_params}
\end{figure*}

\subsubsection{Temperature Profile}

The retrieved P-T profile for each night is shown in Figure~\ref{fig:PT}. In the photosphere at pressures $\sim$1~bar, the agreement between each model is remarkable, with very tight constraints of $\lesssim10$~K in the temperature at 1$\sigma$ confidence thanks to the very high signal-to-noise and spectral resolution of the CRIRES+ observations. In the upper atmosphere, we are less sensitive to the temperature as there are fewer spectral lines which are capable of probing these lower pressures, and hence the uncertainty in the temperature increases as we go up the atmosphere. Similarly, going to deeper pressures below the photosphere also increases the uncertainty in the temperature profile. Figure~\ref{fig:PT} shows that the uncertainty increases rapidly as we go to higher pressures than the cloud deck, as we are no longer sensitive to these deep layers of the atmosphere in the spectrum. There is however a degeneracy between the photospheric temperature gradient just above the cloud deck and the absolute abundances of H$_2$O, $^{12}$CO and $^{13}$CO. A steeper temperature profile such as that for night 1, where the temperature increases more quickly going to higher pressures, results in lower abundances of the chemical species. This is because the data can be fit by having a steeper thermal gradient and lowering the abundances correspondingly to match the line features in the spectrum. On the other hand, the shallower temperature gradient for the other nights require higher abundances. However, we do note that this effect is small and remains within $\sim1\sigma$ for each of the nights.

We also compared our P-T profiles to the best-fit solution from \citet{palma-bifani2023} from Exo-REM self-consistent models \citep{charnay2018}. For pressures $\lesssim$1~bar, the self-consistent model agrees remarkably well with the retrieved P-T profile for each night, $\lesssim$200~K difference from the median best fit for each night. For higher pressures however, the P-T profiles do show differences, as the self-consistent model becomes adiabatic. This may be driven by several factors. Firstly, we are not sensitive to the very deep pressures (P$\gtrsim3$ bar) as they are below the photosphere and below the cloud deck, and hence the uncertainty in the P-T profile increases. However, given our use of the P-T penalty factor for large deviations in the second derivative, the range in P-T profiles in the retrieval may be more limited. Additionally, as we are observing an average over the companion's atmosphere, patchy clouds/inhomogeneities may alter our result compared to the self-consistent model. Nevertheless, the agreement in the regions with the photosphere is encouraging, and the photospheric temperature gradients, which our retrieval is most sensitive to, shows good consistency.

\subsubsection{Clouds}

The cloud deck pressure shows some slight variation between the nights (see Figure~\ref{fig:12_13CO_cloud}). Night 3 indicates the highest pressure for the cloud layer, whereas night 4 indicates the cloud is $\sim$0.2~dex lower pressure. The cloud pressure shows some correlation with the $^{12}$C/$^{13}$C ratio, which is lowest for night 3 and highest for night 4, as previously discussed in section~\ref{sec:isotopes}. This also correlates well with the slightly stronger spectral features in night 3 as seen in Figure~\ref{fig:ccf_med_subtract}. However, the differences in cloud deck constraints are relatively small and within $\sim$1$\sigma$, and overall the atmosphere of AB~Pic~b shows good consistency across the 4 nights of observations. 

We show the retrieved cloud opacity at 2~$\mu$m and 2.5~$\mu$m in Figure~\ref{fig:posterior_clouds}. This shows clearly that there is a preference for a cloud deck whose opacity is decreasing with wavelength in the 2-2.5~$\mu$m range. For all 4 nights the $\kappa_{\mathrm{cl, 2\mu m}}$ is consistently higher than $\kappa_{\mathrm{cl, 2.5\mu m}}$. As the data is high-pass filtered, this difference must be driven by the line feature strengths being altered by the presence of the cloud opacity. Our retrievals with this wavelength-dependent cloud opacity were statistically preferred and gave significantly more reliable results over a model with a single wavelength-independent cloud opacity, which constrained very high CO$_2$ abundances near $\log(\mathrm{CO_2}) \sim -4$ in order to match the continuum. This highlights that with significant wavelength coverage using a grey cloud model may not always be applicable. Generally, cloud decks originating from Fe, Ti and Si condensates are expected to have a drop in opacity in this wavelength range as long as the particle size remains $\lesssim1$~$\mu$m \citep[e.g.][]{wakeford2015, pinhas2017}, consistent with our result. However, note that other chemical species not included in our retrieval with opacity near 2~$\mu$m may also be able to produce this effect.

\subsubsection{Other retrieved parameters}\label{sec:other_params}

Figure~\ref{fig:posterior_other_params} shows the constraints on other retrieval parameters. We find good constraints on the radius of AB~Pic~b. Each of the nights shows a relatively well constrained radius of $\sim$1.85~R$_\mathrm{J}$. However, we note that the radius is degenerate with the photospheric temperature and the data is high-pass filtered to remove the influence of the continuum, which may result in biases occurring on its retrieved value as noted in section~\ref{sec:highpass}). The log(g) shows only a weak constraint near $\log(\mathrm{g/cms^{-2}}) \sim 4.4$. This results in a mass constraint from AB~Pic~b of $25^{+16}_{-9}$~M$_\mathrm{J}$ for night 1, but a higher value with a wider constraint of $\sim 30^{+20}_{-18}$~M$_\mathrm{J}$ for the other nights because the log(g) value is less well constrained for nights 2, 3 and 4. We are only weakly sensitive to the log(g) and it is degenerate with the chemical abundances, and this combined with the uncertainties on the radius makes placing significant constraints on the mass challenging. The mass constraints from each night indicate that AB~Pic~b is more likely a brown dwarf, but with such wide uncertainty it still falls within 1$\sigma$ uncertainty of planetary mass objects. Our constrained mass does appear to be slightly higher than previous studies from evolutionary models which found $\sim$10~M$_\mathrm{J}$ \citep[e.g.][]{bonnefoy2014}, and hence further analyses, potentially without using high-pass filtering on the data, may be required to robustly determine the mass of AB~Pic~b and other companions from atmospheric retrievals.

\begin{figure*}
	\includegraphics[width=\textwidth]{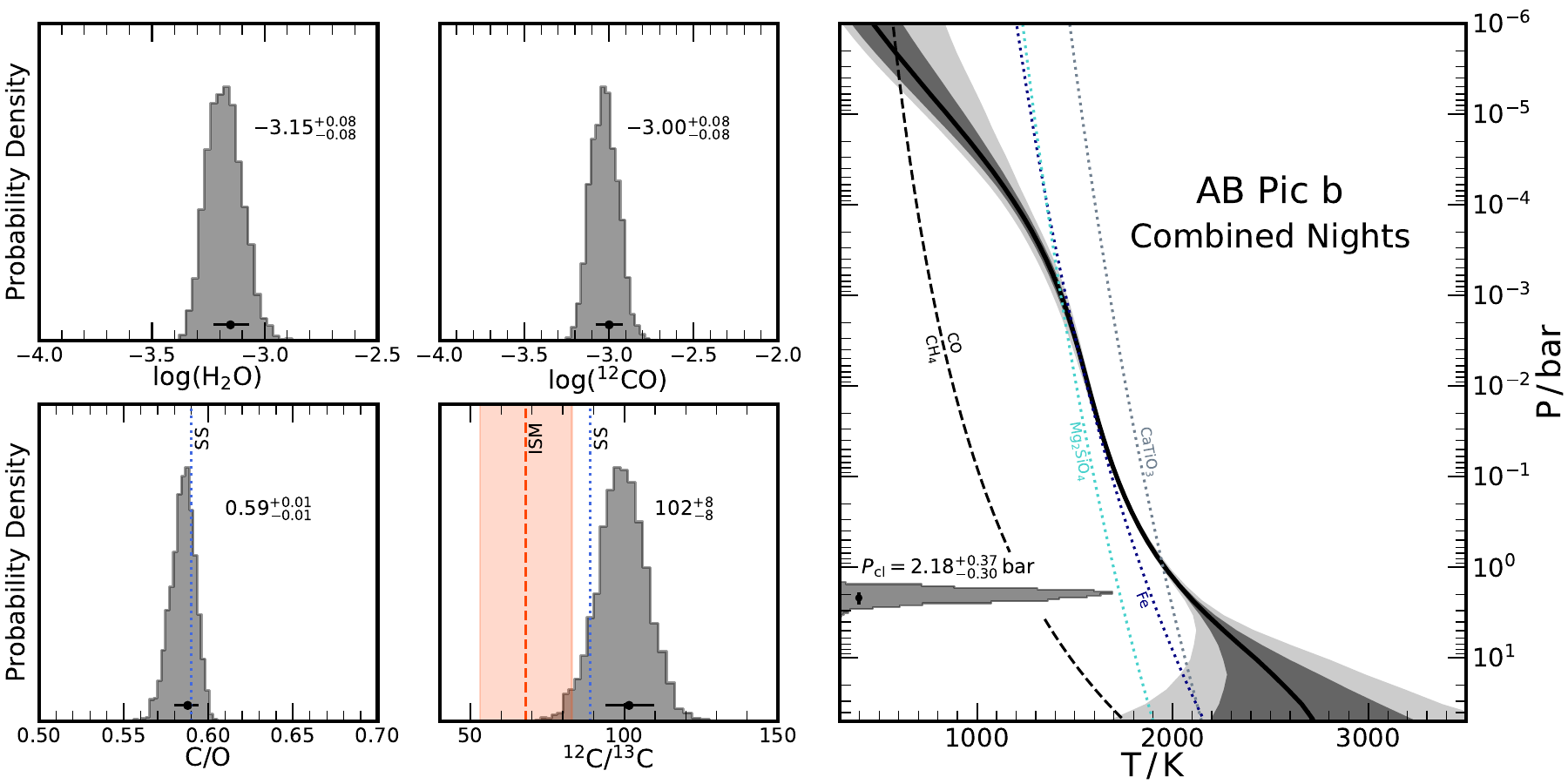}
    \caption{Constraints on the parameters for the combined retrieval of AB~Pic~b simultaneously analysing all four nights. For the isotopic ratio, we show the Solar System value with the blue dotted line and the ISM value and 1$\sigma$ uncertainty with the red line and shaded region \citep{wilson1999, milam2005}, and for the C/O ratio we show the the solar value with the blue dotted line \citep{asplund2021}. The right panel shows the P-T profile and cloud deck constraints, along with the CO-CH$_4$ equilibrium, and the condensation curves for CaTiO$_3$, Fe and Mg$_2$SiO$_4$ \citep{visscher2010, wakeford2017}.}
    \label{fig:combined_nights}
\end{figure*}

We also retrieve the projected rotational velocity ($v\sin i$) of AB~Pic~b. This is achieved by fitting for a rotational broadening kernel on top of the instrumental broadening for each night. This shows that $v\sin i$ is relatively low at $\sim$3.7~km/s. Night 2 does show a higher speed of rotation, and night 1 shows a lower value, but all of them indicate that the spectral lines are not significantly broadened due to the companion's rotation. This gives AB~Pic~b one of the lowest projected rotation velocities measured for companions. This is in contrast to other targets \citep[e.g.][]{snellen2014, deregt2024, landman2024, hsu2024} which showed strong rotational broadening signatures. Such a low $v\sin i$ may be an indicator that AB~Pic~b is either almost pole on as we observe it and/or that it has a low rotation speed due to its young age, as discussed in section~\ref{sec:inclination_obliquity}. While we do see some variability in our retrieved atmospheric parameters ($|\mathrm{CCF}|\lesssim4$), the lack of substantial variability across the 4 nights of observations may indicate that we are observing the companion along its axis of rotation, as the variability is likely to be lower for such a geometry given the same face is always being presented \citep{vos2022}. However, we should note that the measurement of the $v\sin i$ value is highly dependent on the spectral resolution assumed, which in turn is strongly dependent on the seeing given the 0.4" slit used for our observations. Whilst we have accounted for this to determine the resolution for each night, if the actual spectral resolution is higher than that assumed, our $v\sin i$ may also be higher. Previous work by \citet{holmberg2022} report a similar spectral resolution to our nights using the 0.4" slit with AO in the K-band for the LTT-9779 system, indicating that our values are consistent with expectations from the performance of CRIRES+. Other works using the 0.2" slit have also found higher spectral resolutions due to the adaptive optics reducing the point spread function below the slit width \citep[e.g.][]{nortmann2024, pelletier2024}. In addition, the thermal and pressure broadening of spectral lines can also influence the measured value, particularly when the $v\sin i$ is so low.

The Doppler velocity measured for the companion, $\mathrm{RV}$, varies between night 1 and the other nights, with almost 1~km/s difference. This indicates that our wavelength solution may not be well calibrated to the absolute level. The radial velocity of the star, AB~Pic~A, is $\sim22.65$~km/s \citep{soubiran2018}. The four nights do therefore indicate that AB~Pic~b has a line of sight velocity towards us in its orbit, by $\sim1-2$~km/s. This is consistent with a bound orbit around AB~Pic~A, but given that we did not observe AB~Pic~A simultaneously with our observations of AB~Pic~b, it is difficult to further quantify this. Future work, constraining the velocity shift of both the star and the companion, may offer better insight into the radial velocity of AB~Pic~b. Measuring the $\mathrm{RV}$ more accurately also opens up the opportunity to uncover close binary systems, such as the recent discovery of Gliese~229~Ba and Gliese~229~Bb \citep{xuan2024_binary}.

\subsection{Combined Nights}\label{sec:combNights}

\begin{figure*}
	\includegraphics[width=\textwidth]{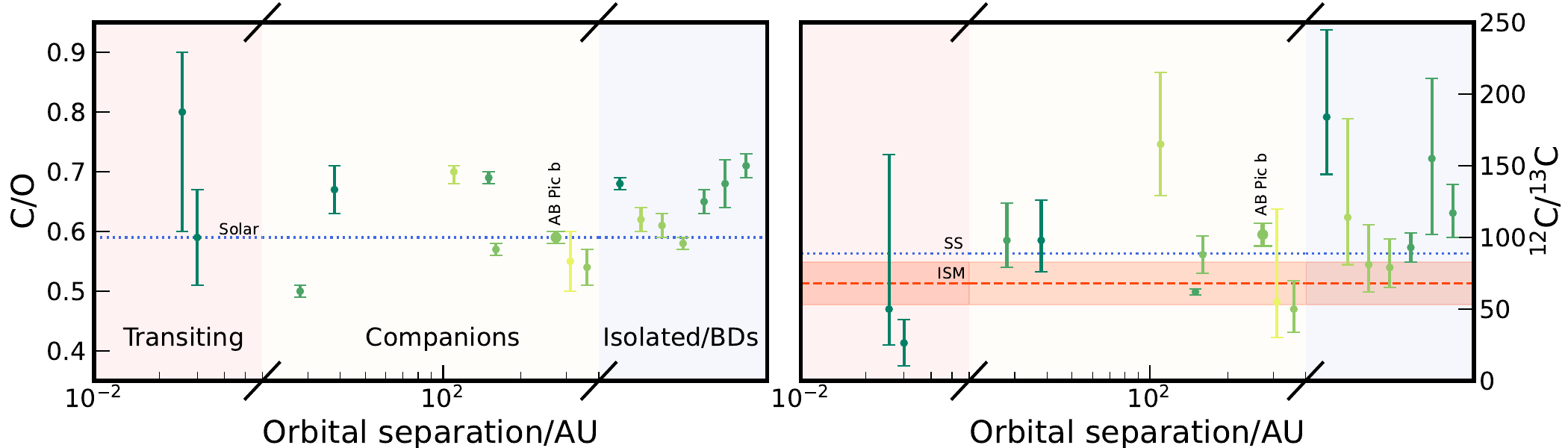}
    \caption{Comparison of our retrieved C/O and $^{12}$C/$^{13}$C ratio from our combined retrieval of all four nights to other close-in planets, companions and isolated targets. These were obtained from \citet{line2021}, \citet{finnerty2023}, \citet{gandhi2023_vhs}, \citet{xuan2024_hip55507}, \citet{costes2024}, \citet{deregt2024}, \citet{gonzalezpicos2024}, \citet{xuan2024}, \citet{zhang2024} and Mulder et al. (2024). The error bars of the sub-stellar objects are colour coded by their age, with yellow being the youngest and dark green being the oldest. We also show the solar C/O ratio \citep{asplund2021} in the left panel with the blue dotted line. In the right panel we show the Solar System $^{12}$C/$^{13}$C value with the blue dotted line and the ISM value and 1$\sigma$ uncertainty with the red dashed line and shaded region respectively \citep{wilson1999, milam2005}.}
    \label{fig:trend_co_iso}
\end{figure*}

Figure~\ref{fig:combined_nights} shows the posterior distributions for the chemical species, the derived C/O and $^{12}$C/$^{13}$C ratios, and the constrained temperature profile for the retrieval combining all 4 nights of observations. The combined retrieval allows for us to determine the effect of analysing all of the observations assuming the same thermal, chemical and planetary structure. However, we still retrieve different GP amplitudes and Doppler velocities for each night. The results are consistent with all parameters near the average of the 4 nights individually. We also obtain tighter constraints on the parameters than each night on its own. the H$_2$O abundance is lower than the $^{12}$CO by $\sim$0.15 dex, indicating that the companion's C/O ratio is $0.59^{+0.01}_{-0.01}$, consistent with the value measured for the Solar System \citep{asplund2021} and the individual nights. The overall abundances of H$_2$O and $^{12}$CO indicate that the atmosphere is slightly above solar metallicity, with [O/H] = 0.54$\pm$0.08 and [C/H] = 0.60$\pm$0.08. Both the overall metallicity and C/O ratio are in excellent agreement with previous works \citep[e.g.][]{bonnefoy2010, palma-bifani2023} using grids of self-consistent models.

The P-T profile for our combined retrieval is shown in the right panel of Figure~\ref{fig:combined_nights}. The profile is in good agreement within $2\sigma$ of the individual nights at all pressures, but there is a slight downward trend of the temperature at the lowest pressures. This may be caused by slight differences in some of the strongest lines in the spectrum, as these probe the lowest pressures and are therefore the most sensitive in the observations. The 0.1~bar photosphere temperature of $1714^{+5}_{-5}$~K is however in excellent agreement with each night and previous works \citep{bonnefoy2010, patience2012, bonnefoy2014, palma-bifani2023}. For almost all pressures our P-T profile is above the CO/CH$_4$ equilibrium boundary, which is consistent with our strong CO constraints but only upper limits on CH$_4$ owing to its lower abundance at these high temperatures.

Our cloud deck pressure is also shown in Figure~\ref{fig:combined_nights} along with the condensation curves for various refractory species at solar abundance \citep{visscher2010, wakeford2017}. We find that our temperature profile constraint most closely matches the CaTiO$_3$ condensation curve at the location of the cloud deck. However, we note that given the semi-grey cloud model we are using, Fe and Mg$_2$SiO$_4$ clouds are also potential sources of our cloud opacity, especially if present at super-solar abundances where they are expected to condense at higher temperatures.

The carbon isotope ratio shows a constraint of $^{12}$C/$^{13}$C~$ = 102^{+8}_{-8}$, in line with the average values obtained for each of the nights individually. This value is slightly higher than that for the Solar System (see Figure~\ref{fig:combined_nights}), but much higher than the value for the ISM. Figure~\ref{fig:trend_co_iso} shows the C/O and isotope ratios for a range of close-in transiting exoplanets, directly-imaged companions and field brown dwarfs. This indicates that AB~Pic~b has a $^{12}$C/$^{13}$C ratio in between that observed for other targets. There appears to be no strong trend over the sample as yet, but future studies exploring isotopic ratios in planetary atmospheres from ground and space may yield insights as to the compositional diversity. We should note however that the individual nights of AB~Pic~b did show some significant differences in the $^{12}$C/$^{13}$C ratio, ranging from values close to the ISM for night 3 to that significantly greater than the Solar System value for night 4. Therefore, the combined retrieval's value should be treated with caution as it is likely affected by atmospheric variability, as discussed in section~\ref{sec:isotopes}.

\subsubsection{Other chemical species}

We find only upper limits for the other chemical species we retrieve. In addition to $^{13}$CO, we retrieved the abundance of C$^{18}$O, which has recently been detected in the atmospheres of directly-imaged companions \citep{gandhi2023_vhs, xuan2024_hip55507}. However, for our retrievals we find $\log(\mathrm{C^{18}O}) \lesssim -6.7$ at 2$\sigma$ confidence. As the $^{16}$O/$^{18}$O abundance ratio is expected to be $\sim500$ \citep{wilson1999}, it is more difficult to detect given its much weaker features compared with $^{13}$CO and hence more observations are needed to constrain this in AB~Pic~b. We also find no clear constraints for CH$_4$, NH$_3$ or CO$_2$ (main isotopologues only), with only an upper limits near $\log(\mathrm{Volume \, Mixing \,Ratio}) \sim -6$ at 2 $\sigma$ confidence. There is a small hint of HCN with a peak in the posterior, which dominated by the data from night 4. However, given that the other nights showed no indication of HCN, we are unable to conclusively detect the species. The CH$_4$ abundance for L-type objects at such high temperatures is expected to be low, and our upper limit is consistent with predictions from chemical modelling \citep{lodders2002, visscher2011}. The CO$_2$ is also relatively unconstrained, as the 2 bluest orders covering $\sim1.9-2.05$~$\mu$m covering strong CO$_2$ absorption were not used in our analysis given the very strong telluric CO$_2$ at those wavelengths. Hence constraining CO$_2$ in exoplanetary atmospheres from ground-based observations may prove more challenging, as noted for H$_2$O \citep[e.g.][]{pelletier2021}. We calculate the C/O ratio of the atmosphere using the total carbon content of the atmosphere divided by the total oxygen content, and so the uncertainty on all of the unconstrained trace species is taken into account when determining the C/O ratio. As the H$_2$O and CO are significantly more abundant than the upper limits for these chemical species, the uncertainty on the unconstrained other species does not significantly alter the C/O ratio constraints for the combined retrieval or for each night individually.

\subsection{Orbit and obliquity}\label{sec:orbital_solution}

Our derived value of $v\sin i$ is much lower than that reported in Table 4 of \citet{palma-bifani2023} ($v\sin i \sim 73$ km/s), used in their Section 5.1 to derive the spin axis of AB~Pic~b. 
The CRIRES+ spectral resolution for each night is much higher than the SINFONI resolution (R $\sim 3000$), allowing us to resolve many more individual lines, and we carefully calibrated the wavelength solution and corrected for instrumental broadening (see Section \ref{sec:combNights}). 
Taking this into account and considering that our $v\sin i$ measurement is stable over the four nights, we decided to re-estimate the spin axis inclination ($i_p$) of AB Pic b. Additionally, a new SPHERE-IRDIS observation was taken in October 2023 (program 0112.C-2045(A); PI P. Nogueira), which we used to refine the orbital fitting to revisit the orbital inclination ($i_o$) of the system. 
We used these two revised angles to gain further insight into the system's obliquity, following the method of \citet{Bryan2021ObliquityB}.

\begin{figure}
	\includegraphics[width=\columnwidth]{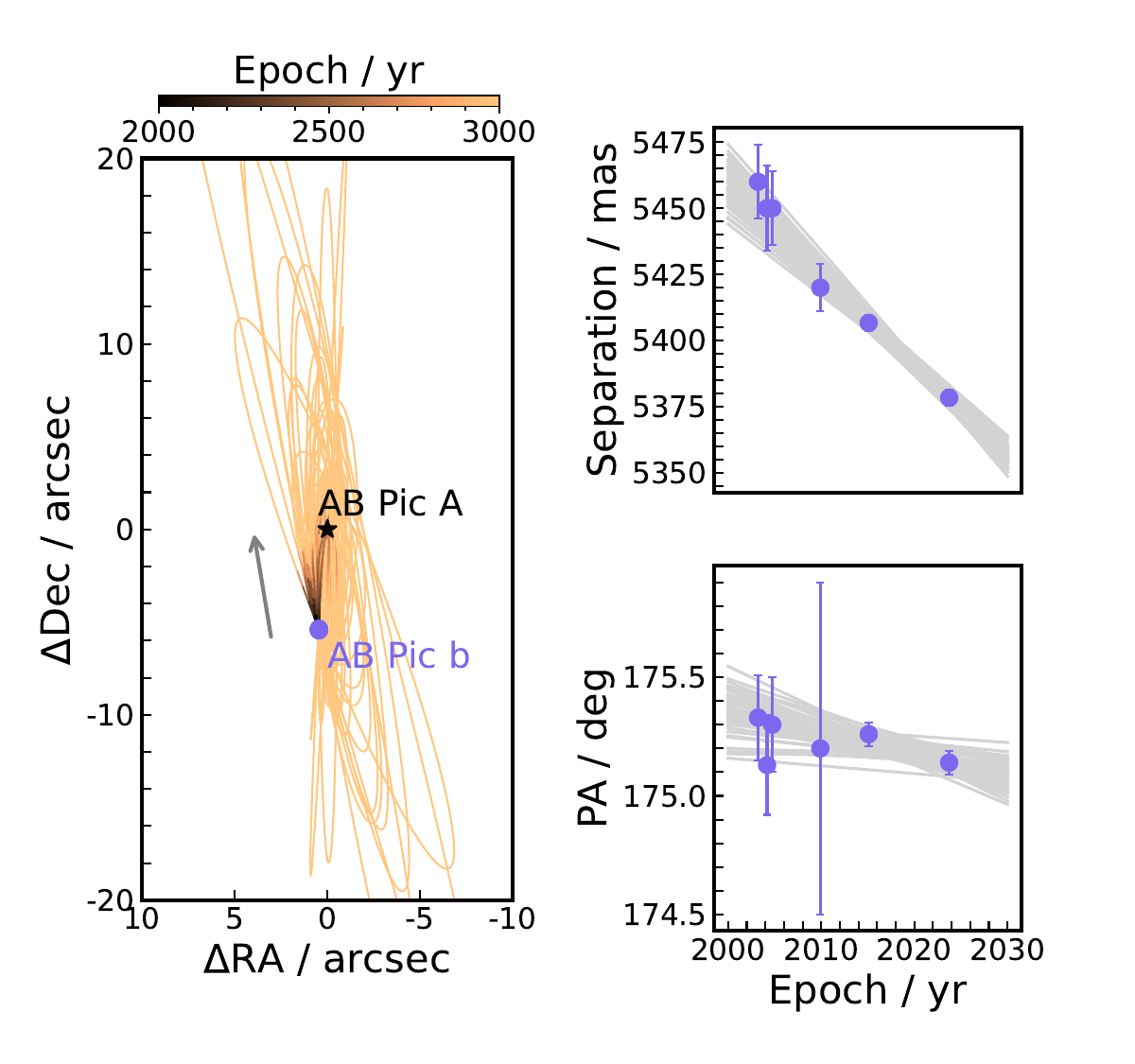}
    \caption{Best orbital solution from \texttt{Orbitize!}. On the left, the positions of AB~Pic~A and AB~Pic~b are marked with a black star and a purple circle, respectively. 
    The colormap represent the predicted position over time for 50 randomly selected orbits, with black indicating earlier epochs and orange later ones. The grey arrow points towards the direction of motion. We marked the position of AB~Pic~b for the six epochs relative to the star, fixed at coordinates (0,0). All purple points overlap since the companion is at a very wide orbit (projected separation $\sim$273 au).
    On the right, we show a zoomed-in view of the astrometry of AB~Pic~b, displaying the separation and position angle as functions of time for the 50 selected orbits.}
    \label{fig:Orbit}
\end{figure}

\subsubsection{Revised orbital solution}

\begin{table}
    \centering
    \resizebox{\columnwidth}{!}{\begin{tabular}{c|c|cc}
\textbf{Epoch} & \textbf{Sep. (mas)}  & \textbf{PA (deg)} & \textbf{Reference}\\
\hline
2003/03/17 & 5460 $\pm$ 14 & 175.33 $\pm$ 0.18 & \citet{chauvin2005} \\ 
2004/03/05 & 5450 $\pm$ 16 & 175.13 $\pm$ 0.21 & \citet{chauvin2005} \\ 
2004/09/25 & 5450 $\pm$ 14 & 175.30 $\pm$ 0.20 & \citet{chauvin2005} \\ 
2009/11/26 & 5420 $\pm$ 9 & 175.2 $\pm$ 0.7 & \citet{rameau2013}\\
2015/02/06  & 5406.69 $\pm$ 3.08 & 175.26 $\pm$ 0.05 & This work\\
2023/10/26 & 5378.39 $\pm$ 2.95 & 175.14 $\pm$ 0.05 & This work\\
\hline
    \end{tabular}}
    \caption{VLT NACO and SPHERE-IRDIS astrometry of AB~Pic~b relative to AB~Pic~A.}
    \label{tab:astrometry}
\end{table}

\begin{figure*}
	\includegraphics[width=\textwidth]{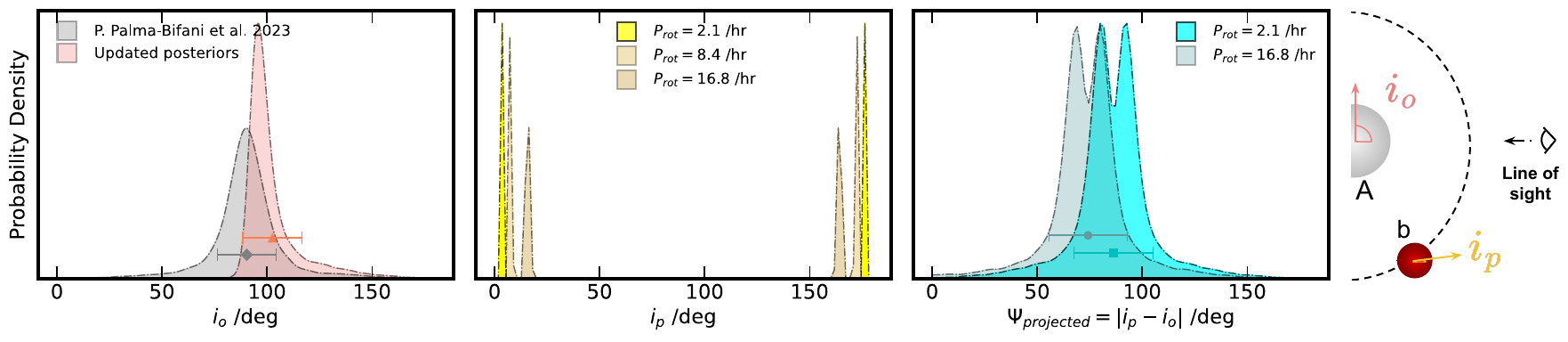}
    \caption{(Left panel) The orbital inclination from \citet{palma-bifani2023} and the updated distribution from this work in grey and light pink respectively. (Center panel) The probability density distributions for the inclination of AB~Pic~b for the reported period of 2.1 hour and for a period 4 and 8 times larger. (Right panel) The revised projected obliquity calculated with a period of 2.1 and 16.8 hours. On the far right, a schematic view of the system is shown, highlighting the inclination vectors of the orbit and the companion.}
    \label{fig:obliquity}
\end{figure*}

We present an updated orbital fit for AB~Pic~b using the \textit{OFTI} (Orbits for the Impatient) module of \textit{orbitize!}, a Bayesian rejection sampling method designed for efficiently fitting the orbits of long-period exoplanets \citep{blunt2017}. For this analysis, we incorporated the data from VLT/NaCo observations in 2003, 2004 \citep{chauvin2005}, and 2009 \citep{rameau2013}, as well as the VLT/SPHERE-IRDIS observations from 2015 already used in \citet{palma-bifani2023} and a new epoch from 2023. We reporcessed for this work both SPHERE-IRDIS epochs using \textit{PACO} \citep{chomez2023} to extract precise astrometry. We report all the astrometry values in Table \ref{tab:astrometry}.
In \textit{orbitize!}, we employed the default priors, as detailed in the documentation, and ran the sampler for 100000 orbits. We draw the best 50 orbital solutions in Figure \ref{fig:Orbit}. The colormap in the left panel of this figure indicates the epoch at which AB~Pic~b is likely to be at the given location with the grey arrow indicating the direction of motion. We can observe that AB~Pic~b is moving towards its host star. To the right, we zoom in at the location of AB~Pic~b and see the measured separation and position angle as a function of time together with the best 50 orbits in grey behind.

Adding the new 2023 VLT/SPHERE-IRDIS data point does not generally impact most posterior distributions, meaning most parameters have the same probability density as reported in \citet{palma-bifani2023}. The impact of this new epoch are a slightly looser constraint on the semi-major axis and a tighter one on the inclination of the orbit. The new semi-major axis is $307^{+305}_{-115}$ AU instead of $190^{+200}_{-50}$ AU, where the errors represent asymmetric 1$\sigma$ confidence intervals assuming Gaussian distributions. 
The new orbital inclination is $98^{+12}_{-5}$ degrees, which suggests a highly inclined orbit but not fully edge-on. We compared the orbital inclination posterior density obtained by \citet{palma-bifani2023} with the one obtained in this work, as shown in the first panel of Figure \ref{fig:obliquity} (in grey and light pink, respectively). In the following section, we use the new inclination distribution for the re-estimation of the obliquity.

\subsubsection{AB~Pic~b inclination and obliquity}\label{sec:inclination_obliquity}

To compute the spin axis of the companion, we use the open-source Python package \textit{exoSpin}\footnote{\url{https://exospin.readthedocs.io/en/latest/}}. This tool calculates the probability density distribution of the spin axis ($i_p$) and obliquity ($\Psi$) of a companion. The method for computing the $i_p$ is based on \citet{bryan2020}, following the same strategy already implemented for AB~Pic~b in \citet{palma-bifani2023}.

We must extract the inclination angle component from the projected rotational velocity ($v\sin i$) measurement to determine the companion's spin axis. For this, we compute the equatorial rotational velocity of AB~Pic~b by assuming solid body rotation, using our derived radius distribution from the combined retrieval and the rotational period reported in \citet{Zhou_2019} of 2.1 hours. 
Since $v$ and $v\sin i$ are not independent, we use Equation 10 from \citet{bryan2020} to derive the probability density distribution for $\cos(i_p)$, from which we obtain the distribution of $i_p$. Our measured $i_p$, reported in the second panel of Figure \ref{fig:obliquity}, is $1.8 \pm 1.3$ degrees, indicating that from our line-of-sight AB~Pic~b is pole-on. This angle is degenerate and symmetric around 90 degrees, so we observe a secondary peak centred at $178.2$ degrees.

Finally, we measure the obliquity from the constraints on the orbital and the companion's inclination. We avoided estimating the true obliquity, as we cannot measure the companion's orientation in a three-dimensional space. In addition, assuming a uniform distribution for the companion's longitude of ascending node ($\Omega$) to compute Equation 11 from \citet{bryan2020} would only increase the uncertainties of our retrieved obliquity. Therefore, we decided to report only the projected obliquity, which is the difference between the inclination vectors for both the companion and orbit, assuming they lie in the same plane. The measured projected obliquity is $86.4 \pm 18.6$ degrees, observable in the right panel of Figure \ref{fig:obliquity} in cyan. Note that this projected obliquity refers to the alignment of the companion's spin axis to the orbital inclination, not to the star's spin axis, as is the case when measuring the Rossiter-McLaughlin effect \citep{Triaud_2018}.

Our low value for the $v\sin i$ may be an indication that the orbit of AB~Pic~b is pole-on. For such a scenario, we would not expect significant modulation of the spectrum that we could attribute to a rotational period (< 0.14 \% for i$_p$ < 3 degrees). This may mean the rotation period is higher than that given in \citet{Zhou_2019}, who find a fast rotation of 2.1 hours at significance level of 1.6 $\sigma$. Such a fast rotation would also mean the companion is very close to its break-up velocity and would be difficult to achieve. Hence we repeated our analysis of i$_p$ for two additional rotational periods of 8.4 and 16.8 hours. We report the results in the second panel of Figure \ref{fig:obliquity}. If AB~Pic~b was rotating with a period of 16.8 hours, the updated value of i$_p$ would be $7.9 \pm 5.6$ degrees, and the updated projected obliquity would be $74.3 \pm 18.5$ degrees. In \citet{palma-bifani2023}, the authors reported a misalignment of the spin axis of AB~Pic~b with respect to the orbital normal axis. A significant misalignment could mean AB~Pic~b is rolling in its orbit and has a Uranus-like orbit and obliquity (see Figure \ref{fig:schematic}), such as that recently reported for VHS~1256~b \citep{poon2024}. 

On the other hand, another potential scenario is that AB~Pic~b is intrinsically a slow rotator. The inclination of the orbit is degenerate if the rotational period is not well known, and hence could exceed 16.8 hours. Companions such as GQ~Lupi~b have been shown to be slow rotators potentially due to their young age \citep{schwarz2016}. AB~Pic~A is a $\sim$13.3~Myr star \citep{booth2021}, and the large radius that we measure for AB~Pic~b could be an indication that the planet has recently formed. Using our measured radius, if the companion's spin axis and orbital inclination are aligned, the orbital period could be as high as $\sim$65~hours. As the planet contracts over time and potentially accretes more material, it may spin up to higher velocities and its period would reduce. Note however that given our use of the high-pass filtering on the observations the constrained radius of AB~Pic~b may not be as reliable, and a smaller radius would result in a lower orbital period. 


\section{Conclusions}

We have observed the directly-imaged companion AB~Pic~b across 4 consecutive nights with the CRIRES+ spectrograph on the VLT. Due to the sensitivity of ground-based high-resolution observations to line depths and shapes and the difficulty in obtaining accurate continuum fluxes, we focus on the variations in the spectral lines to determine variability. The observations showed variations in the line features over the 4 nights, which were relatively small but cross-correlation of the median subtracted data across all nights against a model template was able to detect the differences in the spectra, with $|\mathrm{CCF}|\lesssim4$. We performed atmospheric retrievals across each of the nights individually as well as combined retrievals which simultaneously analysed all four nights together. These retrievals confirmed that the atmosphere of AB~Pic~b showed some variability across the nights, with differences of $2-3\sigma$ uncertainty for a number of parameters.

We constrained the abundance of H$_2$O, $^{12}$CO and $^{13}$CO in the atmosphere of AB~Pic~b, and found that there was some variation in the abundances at the 1-2$\sigma$ level, which was correlated with the temperature gradient at higher pressures. We found a steeper temperature gradient in the deeper atmosphere required a lower abundance of all of the chemical species in order to fit the strength of the observed spectral lines. These differences, while small, do show that constraining absolute abundances from high-resolution spectra is challenging, as seen in other works \citep[e.g.][]{pelletier2023, gandhi2023, maguire2024}.

We also found some variation of the $^{12}$C/$^{13}$C ratio across the nights. Night 4 constrained a value of $^{12}$C/$^{13}$C~$ = 120^{+16}_{-15}$, but night 3 indicated a lower value of $^{12}$C/$^{13}$C~$ = 70^{+16}_{-9}$. This did not seem to be correlated with the detection significances (see Table~\ref{tab:posterior}), but was most strongly correlated with the location of the cloud deck. On night 3 the cloud deck was constrained at deeper pressures than for the other nights. This may indicate that the $^{12}$C/$^{13}$C ratio was lower for this night as the $^{13}$CO features in the spectrum were stronger and required a deeper cloud deck. The $^{13}$CO is less abundant than $^{12}$CO and its spectral features are therefore generated at lower altitudes, hence it is more susceptible to slight variations in the location of the cloud deck as well as the presence of patchy clouds. This shows promise for further studies to investigate and could potentially result in even larger differences in the retrieved $^{12}$C/$^{13}$C ratios between observations if a companion shows significant variability.

We performed a combined retrieval on all four nights simultaneously. This was a way to assess the effect of combining the different datasets which show some variability. Generally, we found that the values for the parameters were the average of the constraints from the individual nights. The H$_2$O and $^{12}$CO abundances were slightly above solar composition, in line with previous studies for similar targets \citep[e.g.][]{xuan2024}. The C/O ratio was consistent with solar composition, but the $^{12}$C/$^{13}$C isotope ratio was slightly higher than that of the Solar System and higher than that seen for the ISM \citep{wilson1999, milam2005}. The isotopic abundance may be used as a way to distinguish formation scenarios \citep{zhang2021}, and does appear to have significant variations for such companions \citep{xuan2024}. Therefore, it is key that we understand the role atmospheric variability plays in determining isotopic ratios. Further observations across a range of Super-Jupiters will also determine whether there are any population level trends across the sample of companions, as well to the sample of field brown dwarfs.

We retrieved the rotational broadening, $v\sin i$, of AB~Pic~b, which we found to be just $\sim$3.7~km/s for the four nights. This allowed us to revisit the obliquity measurement from \citet{palma-bifani2023}. We found that either the $i_p$ is close to 0, indicating that AB~Pic~b is oriented pole-on towards us, and/or that the companion is a slow rotator. The lack of substantial variability in the spectral lines across our four nights of observations may suggest that the companion's rotation axis is aligned with our line of sight, as the same face is always presented. Combining this with the orbital inclination of AB~Pic~b, this may mean that the rotation axis of AB~Pic~b has a $\sim90^\circ$ misalignment to its orbit and hence a Uranus-like orientation. However, for such a geometry the determination of the period would be more challenging, and \citet{Zhou_2019} found a period of $\sim$2.1~hours but at $<2\sigma$ confidence. On the other hand, younger companions have been shown to possess intrinsically lower rotation velocities (and therefore higher periods) \citep{schwarz2016}, and AB~Pic~A is $\sim13.3$~Myr old so the companion could potentially still be forming. The retrieved radius of $\sim$1.9~R$_\mathrm{J}$ may be an indication that this companion has only recently formed and hence the rotation speed could increase over time as it contracts, although it should be noted that our radius value may not be reliable as the data is high-pass filtered. Therefore, a robust determination of the period of AB~Pic~b would resolve these two scenarios and open an exciting avenue for further studies into its dynamic history.

Atmospheric variability has been shown to be a key feature of L and T dwarfs, particularly near the L/T transition \citep[e.g.][]{radigan2014, liu2024}. Younger, lower mass objects have shown even more variation in their spectra over time \citep{vos2022}. Hence, patchy clouds and dynamics are vital to understand and study if we wish to obtain reliable estimates on the chemical composition and other atmospheric properties. High-resolution spectroscopy offers a powerful tool for such studies, enabling Doppler imaging through time-series observations over much shorter timescales \citep{crossfield2014}. In addition, a range of general circulation models have shown how cloud radiative feedback can result in photospheric temperatures varying by $\gtrsim$100~K \citep{tan2021}. In future, higher signal-to-noise observations through instruments such as the ELT's METIS \citep{brandl2021} may help determine the variability of AB~Pic~b and other companions, opening up the exciting opportunity to study the variation and evolution of planetary atmospheres over time.

\section*{Acknowledgements}

The authors would like to acknowledge the University of Warwick Scientific Computing Research Technology Platform (SCRTP) for assistance in the research described in this paper, performed with the Avon HPC cluster.
A.Z. acknowledges support from ANID -- Millennium Science Initiative Program -- Center Code NCN2021\_080. We thank Alex S{\'a}nchez-L{\'o}pez for helpful discussions and feedback. We thank the anonymous referee for a careful review of our manuscript.

\section*{Data Availability}

The data underlying this article are publicly available from the ESO science archive.



\bibliographystyle{mnras}
\bibliography{refs} 

\begin{thebibliography}{}
\makeatletter
\relax
\def\mn@urlcharsother{\let\do\@makeother \do\$\do\&\do\#\do\^\do\_\do\%\do\~}
\def\mn@doi{\begingroup\mn@urlcharsother \@ifnextchar [ {\mn@doi@}
  {\mn@doi@[]}}
\def\mn@doi@[#1]#2{\def\@tempa{#1}\ifx\@tempa\@empty \href
  {http://dx.doi.org/#2} {doi:#2}\else \href {http://dx.doi.org/#2} {#1}\fi
  \endgroup}
\def\mn@eprint#1#2{\mn@eprint@#1:#2::\@nil}
\def\mn@eprint@arXiv#1{\href {http://arxiv.org/abs/#1} {{\tt arXiv:#1}}}
\def\mn@eprint@dblp#1{\href {http://dblp.uni-trier.de/rec/bibtex/#1.xml}
  {dblp:#1}}
\def\mn@eprint@#1:#2:#3:#4\@nil{\def\@tempa {#1}\def\@tempb {#2}\def\@tempc
  {#3}\ifx \@tempc \@empty \let \@tempc \@tempb \let \@tempb \@tempa \fi \ifx
  \@tempb \@empty \def\@tempb {arXiv}\fi \@ifundefined
  {mn@eprint@\@tempb}{\@tempb:\@tempc}{\expandafter \expandafter \csname
  mn@eprint@\@tempb\endcsname \expandafter{\@tempc}}}

\bibitem[\protect\citeauthoryear{{Artigau}, {Bouchard}, {Doyon}  \&
  {Lafreni{\`e}re}}{{Artigau} et~al.}{2009}]{artigau2009}
{Artigau} {\'E}.,  {Bouchard} S.,  {Doyon} R.,   {Lafreni{\`e}re} D.,  2009,
  \mn@doi [\apj] {10.1088/0004-637X/701/2/1534}, \href
  {https://ui.adsabs.harvard.edu/abs/2009ApJ...701.1534A} {701, 1534}

\bibitem[\protect\citeauthoryear{{Asplund}, {Amarsi}  \& {Grevesse}}{{Asplund}
  et~al.}{2021}]{asplund2021}
{Asplund} M.,  {Amarsi} A.~M.,   {Grevesse} N.,  2021, \mn@doi [\aap]
  {10.1051/0004-6361/202140445}, \href
  {https://ui.adsabs.harvard.edu/abs/2021A&A...653A.141A} {653, A141}

\bibitem[\protect\citeauthoryear{{Barber}, {Strange}, {Hill}, {Polyansky},
  {Mellau}, {Yurchenko}  \& {Tennyson}}{{Barber} et~al.}{2014}]{barber2014}
{Barber} R.~J.,  {Strange} J.~K.,  {Hill} C.,  {Polyansky} O.~L.,  {Mellau}
  G.~C.,  {Yurchenko} S.~N.,   {Tennyson} J.,  2014, \mn@doi [Mon. Not. R.
  Astron. Soc.] {10.1093/mnras/stt2011}, \href
  {http://adsabs.harvard.edu/abs/2014MNRAS.437.1828B} {437, 1828}

\bibitem[\protect\citeauthoryear{{Barman}, {Macintosh}, {Konopacky}  \&
  {Marois}}{{Barman} et~al.}{2011}]{barman2011}
{Barman} T.~S.,  {Macintosh} B.,  {Konopacky} Q.~M.,   {Marois} C.,  2011,
  \mn@doi [\apj] {10.1088/0004-637X/733/1/65}, \href
  {https://ui.adsabs.harvard.edu/abs/2011ApJ...733...65B} {733, 65}

\bibitem[\protect\citeauthoryear{{Barrado} et~al.,}{{Barrado}
  et~al.}{2023}]{barrado2023}
{Barrado} D.,  et~al., 2023, \mn@doi [\nat] {10.1038/s41586-023-06813-y}, \href
  {https://ui.adsabs.harvard.edu/abs/2023Natur.624..263B} {624, 263}

\bibitem[\protect\citeauthoryear{{Biller} et~al.,}{{Biller}
  et~al.}{2015}]{biller2015}
{Biller} B.~A.,  et~al., 2015, \mn@doi [\apjl] {10.1088/2041-8205/813/2/L23},
  \href {https://ui.adsabs.harvard.edu/abs/2015ApJ...813L..23B} {813, L23}

\bibitem[\protect\citeauthoryear{{Biller} et~al.,}{{Biller}
  et~al.}{2024}]{biller2024}
{Biller} B.~A.,  et~al., 2024, \mn@doi [\mnras] {10.1093/mnras/stae1602}, \href
  {https://ui.adsabs.harvard.edu/abs/2024MNRAS.532.2207B} {532, 2207}

\bibitem[\protect\citeauthoryear{Blunt et~al.,}{Blunt et~al.}{2017}]{blunt2017}
Blunt S.,  et~al., 2017, \mn@doi [The Astrophysical Journal]
  {10.3847/1538-3881/aa6930}, 153, 229

\bibitem[\protect\citeauthoryear{{Bonnefoy}, {Chauvin}, {Rojo}, {Allard},
  {Lagrange}, {Homeier}, {Dumas}  \& {Beuzit}}{{Bonnefoy}
  et~al.}{2010}]{bonnefoy2010}
{Bonnefoy} M.,  {Chauvin} G.,  {Rojo} P.,  {Allard} F.,  {Lagrange} A.~M.,
  {Homeier} D.,  {Dumas} C.,   {Beuzit} J.~L.,  2010, \mn@doi [\aap]
  {10.1051/0004-6361/200912688}, \href
  {https://ui.adsabs.harvard.edu/abs/2010A&A...512A..52B} {512, A52}

\bibitem[\protect\citeauthoryear{{Bonnefoy}, {Chauvin}, {Lagrange}, {Rojo},
  {Allard}, {Pinte}, {Dumas}  \& {Homeier}}{{Bonnefoy}
  et~al.}{2014}]{bonnefoy2014}
{Bonnefoy} M.,  {Chauvin} G.,  {Lagrange} A.~M.,  {Rojo} P.,  {Allard} F.,
  {Pinte} C.,  {Dumas} C.,   {Homeier} D.,  2014, \mn@doi [\aap]
  {10.1051/0004-6361/201118270}, \href
  {https://ui.adsabs.harvard.edu/abs/2014A&A...562A.127B} {562, A127}

\bibitem[\protect\citeauthoryear{{Booth}, {del Burgo}  \& {Hambaryan}}{{Booth}
  et~al.}{2021}]{booth2021}
{Booth} M.,  {del Burgo} C.,   {Hambaryan} V.~V.,  2021, \mn@doi [\mnras]
  {10.1093/mnras/staa3631}, \href
  {https://ui.adsabs.harvard.edu/abs/2021MNRAS.500.5552B} {500, 5552}

\bibitem[\protect\citeauthoryear{{Brandl} et~al.,}{{Brandl}
  et~al.}{2021}]{brandl2021}
{Brandl} B.,  et~al., 2021, \mn@doi [The Messenger] {10.18727/0722-6691/5218},
  \href {https://ui.adsabs.harvard.edu/abs/2021Msngr.182...22B} {182, 22}

\bibitem[\protect\citeauthoryear{{Bryan}, {Ginzburg}, {Chiang}, {Morley},
  {Bowler}, {Xuan}  \& {Knutson}}{{Bryan} et~al.}{2020}]{bryan2020}
{Bryan} M.~L.,  {Ginzburg} S.,  {Chiang} E.,  {Morley} C.,  {Bowler} B.~P.,
  {Xuan} J.~W.,   {Knutson} H.~A.,  2020, \mn@doi [\apj]
  {10.3847/1538-4357/abc0ef}, \href
  {https://ui.adsabs.harvard.edu/abs/2020ApJ...905...37B} {905, 37}

\bibitem[\protect\citeauthoryear{Bryan, Chiang, Morley, Mace  \& Bowler}{Bryan
  et~al.}{2021}]{Bryan2021ObliquityB}
Bryan M.~L.,  Chiang E.,  Morley C.~V.,  Mace G.~N.,   Bowler B.~P.,  2021,
  Astronomical Journal

\bibitem[\protect\citeauthoryear{{Buchner} et~al.,}{{Buchner}
  et~al.}{2014}]{buchner2014}
{Buchner} J.,  et~al., 2014, \mn@doi [\aap] {10.1051/0004-6361/201322971},
  \href {http://adsabs.harvard.edu/abs/2014A%26A...564A.125B} {564, A125}

\bibitem[\protect\citeauthoryear{{Burningham}, {Marley}, {Line}, {Lupu},
  {Visscher}, {Morley}, {Saumon}  \& {Freedman}}{{Burningham}
  et~al.}{2017}]{burningham2017}
{Burningham} B.,  {Marley} M.~S.,  {Line} M.~R.,  {Lupu} R.,  {Visscher} C.,
  {Morley} C.~V.,  {Saumon} D.,   {Freedman} R.,  2017, \mn@doi [\mnras]
  {10.1093/mnras/stx1246}, \href
  {https://ui.adsabs.harvard.edu/abs/2017MNRAS.470.1177B} {470, 1177}

\bibitem[\protect\citeauthoryear{{Burningham} et~al.,}{{Burningham}
  et~al.}{2021}]{burningham2021}
{Burningham} B.,  et~al., 2021, \mn@doi [\mnras] {10.1093/mnras/stab1361},
  \href {https://ui.adsabs.harvard.edu/abs/2021MNRAS.506.1944B} {506, 1944}

\bibitem[\protect\citeauthoryear{{Charnay}, {B{\'e}zard}, {Baudino},
  {Bonnefoy}, {Boccaletti}  \& {Galicher}}{{Charnay}
  et~al.}{2018}]{charnay2018}
{Charnay} B.,  {B{\'e}zard} B.,  {Baudino} J.~L.,  {Bonnefoy} M.,  {Boccaletti}
  A.,   {Galicher} R.,  2018, \mn@doi [\apj] {10.3847/1538-4357/aaac7d}, \href
  {https://ui.adsabs.harvard.edu/abs/2018ApJ...854..172C} {854, 172}

\bibitem[\protect\citeauthoryear{{Chauvin} et~al.,}{{Chauvin}
  et~al.}{2005}]{chauvin2005}
{Chauvin} G.,  et~al., 2005, \mn@doi [\aap] {10.1051/0004-6361:200500111},
  \href {https://ui.adsabs.harvard.edu/abs/2005A&A...438L..29C} {438, L29}

\bibitem[\protect\citeauthoryear{{Chomez} et~al.,}{{Chomez}
  et~al.}{2023}]{chomez2023}
{Chomez} A.,  et~al., 2023, \mn@doi [\aap] {10.1051/0004-6361/202245723}, \href
  {https://ui.adsabs.harvard.edu/abs/2023A&A...675A.205C} {675, A205}

\bibitem[\protect\citeauthoryear{{Coles}, {Yurchenko}  \& {Tennyson}}{{Coles}
  et~al.}{2019}]{coles2019}
{Coles} P.~A.,  {Yurchenko} S.~N.,   {Tennyson} J.,  2019, \mn@doi [\mnras]
  {10.1093/mnras/stz2778}, \href
  {https://ui.adsabs.harvard.edu/abs/2019MNRAS.490.4638C} {490, 4638}

\bibitem[\protect\citeauthoryear{{Costes} et~al.,}{{Costes}
  et~al.}{2024}]{costes2024}
{Costes} J.~C.,  et~al., 2024, \mn@doi [arXiv e-prints]
  {10.48550/arXiv.2404.11523}, \href
  {https://ui.adsabs.harvard.edu/abs/2024arXiv240411523C} {p. arXiv:2404.11523}

\bibitem[\protect\citeauthoryear{{Crossfield} et~al.,}{{Crossfield}
  et~al.}{2014}]{crossfield2014}
{Crossfield} I.~J.~M.,  et~al., 2014, \mn@doi [\nat] {10.1038/nature12955},
  \href {https://ui.adsabs.harvard.edu/abs/2014Natur.505..654C} {505, 654}

\bibitem[\protect\citeauthoryear{{Currie} et~al.,}{{Currie}
  et~al.}{2011}]{currie2011}
{Currie} T.,  et~al., 2011, \mn@doi [\apj] {10.1088/0004-637X/729/2/128}, \href
  {https://ui.adsabs.harvard.edu/abs/2011ApJ...729..128C} {729, 128}

\bibitem[\protect\citeauthoryear{{Currie}, {Biller}, {Lagrange}, {Marois},
  {Guyon}, {Nielsen}, {Bonnefoy}  \& {De Rosa}}{{Currie}
  et~al.}{2023}]{currie2023}
{Currie} T.,  {Biller} B.,  {Lagrange} A.,  {Marois} C.,  {Guyon} O.,
  {Nielsen} E.~L.,  {Bonnefoy} M.,   {De Rosa} R.~J.,  2023, in {Inutsuka} S.,
  {Aikawa} Y.,  {Muto} T.,  {Tomida} K.,   {Tamura} M.,  eds,  Astronomical
  Society of the Pacific Conference Series Vol. 534, Protostars and Planets
  VII. p.~799 (\mn@eprint {arXiv} {2205.05696}),
  \mn@doi{10.48550/arXiv.2205.05696}

\bibitem[\protect\citeauthoryear{{Cutri} et~al.,}{{Cutri}
  et~al.}{2003}]{cutri2003}
{Cutri} R.~M.,  et~al., 2003, VizieR Online Data Catalog, \href
  {https://ui.adsabs.harvard.edu/abs/2003yCat.2246....0C} {p. II/246}

\bibitem[\protect\citeauthoryear{{Dorn} et~al.,}{{Dorn}
  et~al.}{2014}]{dorn2014}
{Dorn} R.~J.,  et~al., 2014, The Messenger, \href
  {https://ui.adsabs.harvard.edu/abs/2014Msngr.156....7D} {156, 7}

\bibitem[\protect\citeauthoryear{{Dorn} et~al.,}{{Dorn}
  et~al.}{2023}]{dorn2023}
{Dorn} R.~J.,  et~al., 2023, \mn@doi [\aap] {10.1051/0004-6361/202245217},
  \href {https://ui.adsabs.harvard.edu/abs/2023A&A...671A..24D} {671, A24}

\bibitem[\protect\citeauthoryear{{Eriksson}, {Janson}  \&
  {Calissendorff}}{{Eriksson} et~al.}{2019}]{eriksson2019}
{Eriksson} S.~C.,  {Janson} M.,   {Calissendorff} P.,  2019, \mn@doi [\aap]
  {10.1051/0004-6361/201935671}, \href
  {https://ui.adsabs.harvard.edu/abs/2019A&A...629A.145E} {629, A145}

\bibitem[\protect\citeauthoryear{{Feroz} \& {Hobson}}{{Feroz} \&
  {Hobson}}{2008}]{feroz2008}
{Feroz} F.,  {Hobson} M.~P.,  2008, \mn@doi [\mnras]
  {10.1111/j.1365-2966.2007.12353.x}, \href
  {http://adsabs.harvard.edu/abs/2008MNRAS.384..449F} {384, 449}

\bibitem[\protect\citeauthoryear{{Feroz}, {Hobson}  \& {Bridges}}{{Feroz}
  et~al.}{2009}]{feroz2009}
{Feroz} F.,  {Hobson} M.~P.,   {Bridges} M.,  2009, \mn@doi [\mnras]
  {10.1111/j.1365-2966.2009.14548.x}, \href
  {http://adsabs.harvard.edu/abs/2009MNRAS.398.1601F} {398, 1601}

\bibitem[\protect\citeauthoryear{{Feroz}, {Hobson}, {Cameron}  \&
  {Pettitt}}{{Feroz} et~al.}{2013}]{feroz2013}
{Feroz} F.,  {Hobson} M.~P.,  {Cameron} E.,   {Pettitt} A.~N.,  2013, preprint,
  \href {http://adsabs.harvard.edu/abs/2013arXiv1306.2144F} {} (\mn@eprint
  {arXiv} {1306.2144})

\bibitem[\protect\citeauthoryear{{Finnerty} et~al.,}{{Finnerty}
  et~al.}{2023}]{finnerty2023}
{Finnerty} L.,  et~al., 2023, \mn@doi [\aj] {10.3847/1538-3881/acda91}, \href
  {https://ui.adsabs.harvard.edu/abs/2023AJ....166...31F} {166, 31}

\bibitem[\protect\citeauthoryear{{Gaia Collaboration}}{{Gaia
  Collaboration}}{2020}]{gaia2020}
{Gaia Collaboration} 2020, VizieR Online Data Catalog, \href
  {https://ui.adsabs.harvard.edu/abs/2020yCat.1350....0G} {p. I/350}

\bibitem[\protect\citeauthoryear{{Gandhi} \& {Madhusudhan}}{{Gandhi} \&
  {Madhusudhan}}{2017}]{gandhi2017}
{Gandhi} S.,  {Madhusudhan} N.,  2017, \mn@doi [\mnras]
  {10.1093/mnras/stx1601}, \href
  {https://ui.adsabs.harvard.edu/abs/2017MNRAS.472.2334G} {472, 2334}

\bibitem[\protect\citeauthoryear{{Gandhi}, {Madhusudhan}, {Hawker}  \&
  {Piette}}{{Gandhi} et~al.}{2019}]{gandhi2019_hydrah}
{Gandhi} S.,  {Madhusudhan} N.,  {Hawker} G.,   {Piette} A.,  2019, \mn@doi
  [\aj] {10.3847/1538-3881/ab4efc}, \href
  {https://ui.adsabs.harvard.edu/abs/2019AJ....158..228G} {158, 228}

\bibitem[\protect\citeauthoryear{{Gandhi} et~al.,}{{Gandhi}
  et~al.}{2020}]{gandhi2020_cs}
{Gandhi} S.,  et~al., 2020, \mn@doi [\mnras] {10.1093/mnras/staa981}, \href
  {https://ui.adsabs.harvard.edu/abs/2020MNRAS.495..224G} {495, 224}

\bibitem[\protect\citeauthoryear{{Gandhi}, {Kesseli}, {Snellen}, {Brogi},
  {Wardenier}, {Parmentier}, {Welbanks}  \& {Savel}}{{Gandhi}
  et~al.}{2022}]{gandhi2022}
{Gandhi} S.,  {Kesseli} A.,  {Snellen} I.,  {Brogi} M.,  {Wardenier} J.~P.,
  {Parmentier} V.,  {Welbanks} L.,   {Savel} A.~B.,  2022, \mn@doi [\mnras]
  {10.1093/mnras/stac1744}, \href
  {https://ui.adsabs.harvard.edu/abs/2022MNRAS.515..749G} {515, 749}

\bibitem[\protect\citeauthoryear{{Gandhi} et~al.,}{{Gandhi}
  et~al.}{2023a}]{gandhi2023}
{Gandhi} S.,  et~al., 2023a, \mn@doi [\aj] {10.3847/1538-3881/accd65}, \href
  {https://ui.adsabs.harvard.edu/abs/2023AJ....165..242G} {165, 242}

\bibitem[\protect\citeauthoryear{{Gandhi}, {de Regt}, {Snellen}, {Zhang},
  {Rugers}, {van Leur}  \& {Bosschaart}}{{Gandhi}
  et~al.}{2023b}]{gandhi2023_vhs}
{Gandhi} S.,  {de Regt} S.,  {Snellen} I.,  {Zhang} Y.,  {Rugers} B.,  {van
  Leur} N.,   {Bosschaart} Q.,  2023b, \mn@doi [\apjl]
  {10.3847/2041-8213/ad07e2}, \href
  {https://ui.adsabs.harvard.edu/abs/2023ApJ...957L..36G} {957, L36}

\bibitem[\protect\citeauthoryear{{Gibson} et~al.,}{{Gibson}
  et~al.}{2020}]{gibson2020}
{Gibson} N.~P.,  et~al., 2020, \mn@doi [\mnras] {10.1093/mnras/staa228}, \href
  {https://ui.adsabs.harvard.edu/abs/2020MNRAS.493.2215G} {493, 2215}

\bibitem[\protect\citeauthoryear{{Gonz{\'a}lez Picos} et~al.,}{{Gonz{\'a}lez
  Picos} et~al.}{2024}]{gonzalezpicos2024}
{Gonz{\'a}lez Picos} D.,  et~al., 2024, \mn@doi [arXiv e-prints]
  {10.48550/arXiv.2407.07678}, \href
  {https://ui.adsabs.harvard.edu/abs/2024arXiv240707678G} {p. arXiv:2407.07678}

\bibitem[\protect\citeauthoryear{{Hallinan} et~al.,}{{Hallinan}
  et~al.}{2015}]{hallinan2015}
{Hallinan} G.,  et~al., 2015, \mn@doi [\nat] {10.1038/nature14619}, \href
  {https://ui.adsabs.harvard.edu/abs/2015Natur.523..568H} {523, 568}

\bibitem[\protect\citeauthoryear{{Hargreaves}, {Gordon}, {Rey}, {Nikitin},
  {Tyuterev}, {Kochanov}  \& {Rothman}}{{Hargreaves}
  et~al.}{2020}]{hargreaves2020}
{Hargreaves} R.~J.,  {Gordon} I.~E.,  {Rey} M.,  {Nikitin} A.~V.,  {Tyuterev}
  V.~G.,  {Kochanov} R.~V.,   {Rothman} L.~S.,  2020, \mn@doi [\apjs]
  {10.3847/1538-4365/ab7a1a}, \href
  {https://ui.adsabs.harvard.edu/abs/2020ApJS..247...55H} {247, 55}

\bibitem[\protect\citeauthoryear{{Harris}, {Tennyson}, {Kaminsky}, {Pavlenko}
  \& {Jones}}{{Harris} et~al.}{2006}]{harris2006}
{Harris} G.~J.,  {Tennyson} J.,  {Kaminsky} B.~M.,  {Pavlenko} Y.~V.,   {Jones}
  H.~R.~A.,  2006, \mn@doi [Mon. Not. R. Astron. Soc.]
  {10.1111/j.1365-2966.2005.09960.x}, \href
  {http://adsabs.harvard.edu/abs/2006MNRAS.367..400H} {367, 400}

\bibitem[\protect\citeauthoryear{{Holmberg} \& {Madhusudhan}}{{Holmberg} \&
  {Madhusudhan}}{2022}]{holmberg2022}
{Holmberg} M.,  {Madhusudhan} N.,  2022, \mn@doi [\aj]
  {10.3847/1538-3881/ac77eb}, \href
  {https://ui.adsabs.harvard.edu/abs/2022AJ....164...79H} {164, 79}

\bibitem[\protect\citeauthoryear{{Hood} et~al.,}{{Hood}
  et~al.}{2024}]{hood2024}
{Hood} C.~E.,  et~al., 2024, \mn@doi [arXiv e-prints]
  {10.48550/arXiv.2402.05345}, \href
  {https://ui.adsabs.harvard.edu/abs/2024arXiv240205345H} {p. arXiv:2402.05345}

\bibitem[\protect\citeauthoryear{{Horne}}{{Horne}}{1986}]{horne1986}
{Horne} K.,  1986, \mn@doi [\pasp] {10.1086/131801}, \href
  {https://ui.adsabs.harvard.edu/abs/1986PASP...98..609H} {98, 609}

\bibitem[\protect\citeauthoryear{{Hsu} et~al.,}{{Hsu} et~al.}{2024}]{hsu2024}
{Hsu} C.-C.,  et~al., 2024, \mn@doi [arXiv e-prints]
  {10.48550/arXiv.2405.08312}, \href
  {https://ui.adsabs.harvard.edu/abs/2024arXiv240508312H} {p. arXiv:2405.08312}

\bibitem[\protect\citeauthoryear{{Huang}, {Freedman}, {Tashkun}, {Schwenke}  \&
  {Lee}}{{Huang} et~al.}{2013}]{huang2013}
{Huang} X.,  {Freedman} R.~S.,  {Tashkun} S.~A.,  {Schwenke} D.~W.,   {Lee}
  T.~J.,  2013, \mn@doi [\jqsrt] {10.1016/j.jqsrt.2013.05.018}, \href
  {https://ui.adsabs.harvard.edu/abs/2013JQSRT.130..134H} {130, 134}

\bibitem[\protect\citeauthoryear{Huang, Schwenke, Freedman  \& Lee}{Huang
  et~al.}{2017}]{huang2017}
Huang X.,  Schwenke D.~W.,  Freedman R.~S.,   Lee T.~J.,  2017, \mn@doi
  [Journal of Quantitative Spectroscopy and Radiative Transfer]
  {https://doi.org/10.1016/j.jqsrt.2017.04.026}, 203, 224

\bibitem[\protect\citeauthoryear{{Jones}, {Noll}, {Kausch}, {Szyszka}  \&
  {Kimeswenger}}{{Jones} et~al.}{2013}]{jones2013}
{Jones} A.,  {Noll} S.,  {Kausch} W.,  {Szyszka} C.,   {Kimeswenger} S.,  2013,
  \mn@doi [\aap] {10.1051/0004-6361/201322433}, \href
  {https://ui.adsabs.harvard.edu/abs/2013A&A...560A..91J} {560, A91}

\bibitem[\protect\citeauthoryear{{Kaeufl} et~al.,}{{Kaeufl}
  et~al.}{2004}]{kaeufl2004}
{Kaeufl} H.-U.,  et~al., 2004, in {Moorwood} A. F.~M.,  {Iye} M.,  eds,
  Society of Photo-Optical Instrumentation Engineers (SPIE) Conference Series
  Vol. 5492, Ground-based Instrumentation for Astronomy. pp 1218--1227,
  \mn@doi{10.1117/12.551480}

\bibitem[\protect\citeauthoryear{{Landman} et~al.,}{{Landman}
  et~al.}{2024}]{landman2024}
{Landman} R.,  et~al., 2024, \mn@doi [\aap] {10.1051/0004-6361/202347846},
  \href {https://ui.adsabs.harvard.edu/abs/2024A&A...682A..48L} {682, A48}

\bibitem[\protect\citeauthoryear{{Lavie} et~al.,}{{Lavie}
  et~al.}{2017}]{lavie2017}
{Lavie} B.,  et~al., 2017, \mn@doi [\aj] {10.3847/1538-3881/aa7ed8}, \href
  {https://ui.adsabs.harvard.edu/abs/2017AJ....154...91L} {154, 91}

\bibitem[\protect\citeauthoryear{{Lee}, {Tan}  \& {Tsai}}{{Lee}
  et~al.}{2024}]{lee2024}
{Lee} E. K.~H.,  {Tan} X.,   {Tsai} S.-M.,  2024, \mn@doi [\mnras]
  {10.1093/mnras/stae537}, \href
  {https://ui.adsabs.harvard.edu/abs/2024MNRAS.529.2686L} {529, 2686}

\bibitem[\protect\citeauthoryear{{Lew} et~al.,}{{Lew} et~al.}{2024}]{lew2024}
{Lew} B. W.~P.,  et~al., 2024, \mn@doi [\aj] {10.3847/1538-3881/ad3425}, \href
  {https://ui.adsabs.harvard.edu/abs/2024AJ....167..237L} {167, 237}

\bibitem[\protect\citeauthoryear{{Li} \& {Cao}}{{Li} \& {Cao}}{2022}]{li2022}
{Li} Z.,  {Cao} J.,  2022, \mn@doi [arXiv e-prints]
  {10.48550/arXiv.2201.06808}, \href
  {https://ui.adsabs.harvard.edu/abs/2022arXiv220106808L} {p. arXiv:2201.06808}

\bibitem[\protect\citeauthoryear{Li, Gordon, Rothman, Tan, Hu, Kassi, Campargue
   \& Medvedev}{Li et~al.}{2015}]{li2015}
Li G.,  Gordon I.~E.,  Rothman L.~S.,  Tan Y.,  Hu S.-M.,  Kassi S.,  Campargue
  A.,   Medvedev E.~S.,  2015, \mn@doi [The Astrophysical Journal Supplement
  Series] {10.1088/0067-0049/216/1/15}, 216, 15

\bibitem[\protect\citeauthoryear{{Line}, {Teske}, {Burningham}, {Fortney}  \&
  {Marley}}{{Line} et~al.}{2015}]{line2015}
{Line} M.~R.,  {Teske} J.,  {Burningham} B.,  {Fortney} J.~J.,   {Marley}
  M.~S.,  2015, \mn@doi [\apj] {10.1088/0004-637X/807/2/183}, \href
  {https://ui.adsabs.harvard.edu/abs/2015ApJ...807..183L} {807, 183}

\bibitem[\protect\citeauthoryear{{Line} et~al.,}{{Line}
  et~al.}{2017}]{line2017}
{Line} M.~R.,  et~al., 2017, \mn@doi [\apj] {10.3847/1538-4357/aa7ff0}, \href
  {https://ui.adsabs.harvard.edu/abs/2017ApJ...848...83L} {848, 83}

\bibitem[\protect\citeauthoryear{{Line} et~al.,}{{Line}
  et~al.}{2021}]{line2021}
{Line} M.~R.,  et~al., 2021, \mn@doi [\nat] {10.1038/s41586-021-03912-6}, \href
  {https://ui.adsabs.harvard.edu/abs/2021Natur.598..580L} {598, 580}

\bibitem[\protect\citeauthoryear{{Liu} et~al.,}{{Liu} et~al.}{2024}]{liu2024}
{Liu} P.,  et~al., 2024, \mn@doi [\mnras] {10.1093/mnras/stad3502}, \href
  {https://ui.adsabs.harvard.edu/abs/2024MNRAS.527.6624L} {527, 6624}

\bibitem[\protect\citeauthoryear{{Lodders} \& {Fegley}}{{Lodders} \&
  {Fegley}}{2002}]{lodders2002}
{Lodders} K.,  {Fegley} B.,  2002, \mn@doi [\icarus] {10.1006/icar.2001.6740},
  \href {https://ui.adsabs.harvard.edu/abs/2002Icar..155..393L} {155, 393}

\bibitem[\protect\citeauthoryear{{Madhusudhan}}{{Madhusudhan}}{2012}]{madhu2012}
{Madhusudhan} N.,  2012, \mn@doi [\apj] {10.1088/0004-637X/758/1/36}, \href
  {https://ui.adsabs.harvard.edu/abs/2012ApJ...758...36M} {758, 36}

\bibitem[\protect\citeauthoryear{{Maguire}, {Gibson}, {Nugroho}, {Fortune},
  {Ramkumar}, {Gandhi}  \& {de Mooij}}{{Maguire} et~al.}{2024}]{maguire2024}
{Maguire} C.,  {Gibson} N.~P.,  {Nugroho} S.~K.,  {Fortune} M.,  {Ramkumar} S.,
   {Gandhi} S.,   {de Mooij} E.,  2024, \mn@doi [\aap]
  {10.1051/0004-6361/202449449}, \href
  {https://ui.adsabs.harvard.edu/abs/2024A&A...687A..49M} {687, A49}

\bibitem[\protect\citeauthoryear{{Marley}, {Saumon}, {Cushing}, {Ackerman},
  {Fortney}  \& {Freedman}}{{Marley} et~al.}{2012}]{marley2012}
{Marley} M.~S.,  {Saumon} D.,  {Cushing} M.,  {Ackerman} A.~S.,  {Fortney}
  J.~J.,   {Freedman} R.,  2012, \mn@doi [\apj] {10.1088/0004-637X/754/2/135},
  \href {https://ui.adsabs.harvard.edu/abs/2012ApJ...754..135M} {754, 135}

\bibitem[\protect\citeauthoryear{{McCarthy} et~al.,}{{McCarthy}
  et~al.}{2024}]{mccarthy2024}
{McCarthy} A.~M.,  et~al., 2024, arXiv e-prints, \href
  {https://ui.adsabs.harvard.edu/abs/2024arXiv241116577M} {p. arXiv:2411.16577}

\bibitem[\protect\citeauthoryear{{Metchev} et~al.,}{{Metchev}
  et~al.}{2015}]{metchev2015}
{Metchev} S.~A.,  et~al., 2015, \mn@doi [\apj] {10.1088/0004-637X/799/2/154},
  \href {https://ui.adsabs.harvard.edu/abs/2015ApJ...799..154M} {799, 154}

\bibitem[\protect\citeauthoryear{{Milam}, {Savage}, {Brewster}, {Ziurys}  \&
  {Wyckoff}}{{Milam} et~al.}{2005}]{milam2005}
{Milam} S.~N.,  {Savage} C.,  {Brewster} M.~A.,  {Ziurys} L.~M.,   {Wyckoff}
  S.,  2005, \mn@doi [\apj] {10.1086/497123}, \href
  {https://ui.adsabs.harvard.edu/abs/2005ApJ...634.1126M} {634, 1126}

\bibitem[\protect\citeauthoryear{{Miotello}, {Bruderer}  \& {van
  Dishoeck}}{{Miotello} et~al.}{2014}]{miotello2014}
{Miotello} A.,  {Bruderer} S.,   {van Dishoeck} E.~F.,  2014, \mn@doi [\aap]
  {10.1051/0004-6361/201424712}, \href
  {https://ui.adsabs.harvard.edu/abs/2014A&A...572A..96M} {572, A96}

\bibitem[\protect\citeauthoryear{{Molli{\`e}re} \& {Snellen}}{{Molli{\`e}re} \&
  {Snellen}}{2019}]{molliere2019}
{Molli{\`e}re} P.,  {Snellen} I.~A.~G.,  2019, \mn@doi [\aap]
  {10.1051/0004-6361/201834169}, \href
  {https://ui.adsabs.harvard.edu/abs/2019A&A...622A.139M} {622, A139}

\bibitem[\protect\citeauthoryear{{Molli{\`e}re} et~al.,}{{Molli{\`e}re}
  et~al.}{2020}]{molliere2020}
{Molli{\`e}re} P.,  et~al., 2020, \mn@doi [\aap] {10.1051/0004-6361/202038325},
  \href {https://ui.adsabs.harvard.edu/abs/2020A&A...640A.131M} {640, A131}

\bibitem[\protect\citeauthoryear{{Mordasini}, {van Boekel}, {Molli{\`e}re},
  {Henning}  \& {Benneke}}{{Mordasini} et~al.}{2016}]{mordasini2016}
{Mordasini} C.,  {van Boekel} R.,  {Molli{\`e}re} P.,  {Henning} T.,
  {Benneke} B.,  2016, \mn@doi [\apj] {10.3847/0004-637X/832/1/41}, \href
  {https://ui.adsabs.harvard.edu/abs/2016ApJ...832...41M} {832, 41}

\bibitem[\protect\citeauthoryear{{Morley}, {Skemer}, {Miles}, {Line}, {Lopez},
  {Brogi}, {Freedman}  \& {Marley}}{{Morley} et~al.}{2019}]{morley2019}
{Morley} C.~V.,  {Skemer} A.~J.,  {Miles} B.~E.,  {Line} M.~R.,  {Lopez} E.~D.,
   {Brogi} M.,  {Freedman} R.~S.,   {Marley} M.~S.,  2019, \mn@doi [\apjl]
  {10.3847/2041-8213/ab3c65}, \href
  {https://ui.adsabs.harvard.edu/abs/2019ApJ...882L..29M} {882, L29}

\bibitem[\protect\citeauthoryear{{Morris} et~al.,}{{Morris}
  et~al.}{2024}]{morris2024}
{Morris} E.~C.,  et~al., 2024, \mn@doi [\aj] {10.3847/1538-3881/ad4ecf}, \href
  {https://ui.adsabs.harvard.edu/abs/2024AJ....168..144M} {168, 144}

\bibitem[\protect\citeauthoryear{{Noll}, {Kausch}, {Barden}, {Jones},
  {Szyszka}, {Kimeswenger}  \& {Vinther}}{{Noll} et~al.}{2012}]{noll2012}
{Noll} S.,  {Kausch} W.,  {Barden} M.,  {Jones} A.~M.,  {Szyszka} C.,
  {Kimeswenger} S.,   {Vinther} J.,  2012, \mn@doi [\aap]
  {10.1051/0004-6361/201219040}, \href
  {https://ui.adsabs.harvard.edu/abs/2012A&A...543A..92N} {543, A92}

\bibitem[\protect\citeauthoryear{{Nortmann} et~al.,}{{Nortmann}
  et~al.}{2024}]{nortmann2024}
{Nortmann} L.,  et~al., 2024, \mn@doi [arXiv e-prints]
  {10.48550/arXiv.2404.12363}, \href
  {https://ui.adsabs.harvard.edu/abs/2024arXiv240412363N} {p. arXiv:2404.12363}

\bibitem[\protect\citeauthoryear{{{\"O}berg}, {Murray-Clay}  \&
  {Bergin}}{{{\"O}berg} et~al.}{2011}]{oberg2011}
{{\"O}berg} K.~I.,  {Murray-Clay} R.,   {Bergin} E.~A.,  2011, \mn@doi [\apjl]
  {10.1088/2041-8205/743/1/L16}, \href
  {https://ui.adsabs.harvard.edu/abs/2011ApJ...743L..16O} {743, L16}

\bibitem[\protect\citeauthoryear{{Palma-Bifani} et~al.,}{{Palma-Bifani}
  et~al.}{2023}]{palma-bifani2023}
{Palma-Bifani} P.,  et~al., 2023, \mn@doi [\aap] {10.1051/0004-6361/202244294},
  \href {https://ui.adsabs.harvard.edu/abs/2023A&A...670A..90P} {670, A90}

\bibitem[\protect\citeauthoryear{{Patience}, {King}, {De Rosa}, {Vigan},
  {Witte}, {Rice}, {Helling}  \& {Hauschildt}}{{Patience}
  et~al.}{2012}]{patience2012}
{Patience} J.,  {King} R.~R.,  {De Rosa} R.~J.,  {Vigan} A.,  {Witte} S.,
  {Rice} E.,  {Helling} C.,   {Hauschildt} P.,  2012, \mn@doi [\aap]
  {10.1051/0004-6361/201118058}, \href
  {https://ui.adsabs.harvard.edu/abs/2012A&A...540A..85P} {540, A85}

\bibitem[\protect\citeauthoryear{{Pelletier} et~al.,}{{Pelletier}
  et~al.}{2021}]{pelletier2021}
{Pelletier} S.,  et~al., 2021, \mn@doi [\aj] {10.3847/1538-3881/ac0428}, \href
  {https://ui.adsabs.harvard.edu/abs/2021AJ....162...73P} {162, 73}

\bibitem[\protect\citeauthoryear{{Pelletier} et~al.,}{{Pelletier}
  et~al.}{2023}]{pelletier2023}
{Pelletier} S.,  et~al., 2023, \mn@doi [\nat] {10.1038/s41586-023-06134-0},
  \href {https://ui.adsabs.harvard.edu/abs/2023Natur.619..491P} {619, 491}

\bibitem[\protect\citeauthoryear{{Pelletier} et~al.,}{{Pelletier}
  et~al.}{2024}]{pelletier2024}
{Pelletier} S.,  et~al., 2024, \mn@doi [arXiv e-prints]
  {10.48550/arXiv.2410.18183}, \href
  {https://ui.adsabs.harvard.edu/abs/2024arXiv241018183P} {p. arXiv:2410.18183}

\bibitem[\protect\citeauthoryear{{Pinhas} \& {Madhusudhan}}{{Pinhas} \&
  {Madhusudhan}}{2017}]{pinhas2017}
{Pinhas} A.,  {Madhusudhan} N.,  2017, \mn@doi [\mnras]
  {10.1093/mnras/stx1849}, \href
  {https://ui.adsabs.harvard.edu/abs/2017MNRAS.471.4355P} {471, 4355}

\bibitem[\protect\citeauthoryear{{Polyansky}, {Kyuberis}, {Zobov}, {Tennyson},
  {Yurchenko}  \& {Lodi}}{{Polyansky} et~al.}{2018}]{polyansky2018}
{Polyansky} O.~L.,  {Kyuberis} A.~A.,  {Zobov} N.~F.,  {Tennyson} J.,
  {Yurchenko} S.~N.,   {Lodi} L.,  2018, \mn@doi [\mnras]
  {10.1093/mnras/sty1877}, \href
  {https://ui.adsabs.harvard.edu/abs/2018MNRAS.480.2597P} {480, 2597}

\bibitem[\protect\citeauthoryear{{Poon}, {Bryan}, {Rein}, {Morley}, {Mace},
  {Zhou}  \& {Bowler}}{{Poon} et~al.}{2024}]{poon2024}
{Poon} M.,  {Bryan} M.,  {Rein} H.,  {Morley} C.,  {Mace} G.,  {Zhou} Y.,
  {Bowler} B.,  2024, in AAS/Division of Dynamical Astronomy Meeting. p. 100.04

\bibitem[\protect\citeauthoryear{{Radigan}, {Jayawardhana}, {Lafreni{\`e}re},
  {Artigau}, {Marley}  \& {Saumon}}{{Radigan} et~al.}{2012}]{radigan2012}
{Radigan} J.,  {Jayawardhana} R.,  {Lafreni{\`e}re} D.,  {Artigau} {\'E}.,
  {Marley} M.,   {Saumon} D.,  2012, \mn@doi [\apj]
  {10.1088/0004-637X/750/2/105}, \href
  {https://ui.adsabs.harvard.edu/abs/2012ApJ...750..105R} {750, 105}

\bibitem[\protect\citeauthoryear{{Radigan}, {Lafreni{\`e}re}, {Jayawardhana}
  \& {Artigau}}{{Radigan} et~al.}{2014}]{radigan2014}
{Radigan} J.,  {Lafreni{\`e}re} D.,  {Jayawardhana} R.,   {Artigau} E.,  2014,
  \mn@doi [\apj] {10.1088/0004-637X/793/2/75}, \href
  {https://ui.adsabs.harvard.edu/abs/2014ApJ...793...75R} {793, 75}

\bibitem[\protect\citeauthoryear{{Rameau} et~al.,}{{Rameau}
  et~al.}{2013}]{rameau2013}
{Rameau} J.,  et~al., 2013, \mn@doi [\aap] {10.1051/0004-6361/201220984}, \href
  {https://ui.adsabs.harvard.edu/abs/2013A&A...553A..60R} {553, A60}

\bibitem[\protect\citeauthoryear{{Richard} et~al.,}{{Richard}
  et~al.}{2012}]{richard2012}
{Richard} C.,  et~al., 2012, \mn@doi [\jqsrt] {10.1016/j.jqsrt.2011.11.004},
  \href {https://ui.adsabs.harvard.edu/abs/2012JQSRT.113.1276R} {113, 1276}

\bibitem[\protect\citeauthoryear{{Robinson} \& {Marley}}{{Robinson} \&
  {Marley}}{2014}]{robinson2014}
{Robinson} T.~D.,  {Marley} M.~S.,  2014, \mn@doi [\apj]
  {10.1088/0004-637X/785/2/158}, \href
  {https://ui.adsabs.harvard.edu/abs/2014ApJ...785..158R} {785, 158}

\bibitem[\protect\citeauthoryear{{Rothman} et~al.,}{{Rothman}
  et~al.}{2010}]{rothman2010}
{Rothman} L.~S.,  et~al., 2010, \mn@doi [JQSRT] {10.1016/j.jqsrt.2010.05.001},
  \href {http://adsabs.harvard.edu/abs/2010JQSRT.111.2139R} {111, 2139}

\bibitem[\protect\citeauthoryear{{Rowland} et~al.,}{{Rowland}
  et~al.}{2024}]{rowland2024}
{Rowland} M.~J.,  et~al., 2024, \mn@doi [arXiv e-prints]
  {10.48550/arXiv.2411.14541}, \href
  {https://ui.adsabs.harvard.edu/abs/2024arXiv241114541R} {p. arXiv:2411.14541}

\bibitem[\protect\citeauthoryear{{Ruffio} et~al.,}{{Ruffio}
  et~al.}{2019}]{ruffio2019}
{Ruffio} J.-B.,  et~al., 2019, \mn@doi [\aj] {10.3847/1538-3881/ab4594}, \href
  {https://ui.adsabs.harvard.edu/abs/2019AJ....158..200R} {158, 200}

\bibitem[\protect\citeauthoryear{{Schwarz}, {Ginski}, {de Kok}, {Snellen},
  {Brogi}  \& {Birkby}}{{Schwarz} et~al.}{2016}]{schwarz2016}
{Schwarz} H.,  {Ginski} C.,  {de Kok} R.~J.,  {Snellen} I. A.~G.,  {Brogi} M.,
   {Birkby} J.~L.,  2016, \mn@doi [\aap] {10.1051/0004-6361/201628908}, \href
  {https://ui.adsabs.harvard.edu/abs/2016A&A...593A..74S} {593, A74}

\bibitem[\protect\citeauthoryear{{Snellen}, {Brandl}, {de Kok}, {Brogi},
  {Birkby}  \& {Schwarz}}{{Snellen} et~al.}{2014}]{snellen2014}
{Snellen} I. A.~G.,  {Brandl} B.~R.,  {de Kok} R.~J.,  {Brogi} M.,  {Birkby}
  J.,   {Schwarz} H.,  2014, \mn@doi [\nat] {10.1038/nature13253}, \href
  {https://ui.adsabs.harvard.edu/abs/2014Natur.509...63S} {509, 63}

\bibitem[\protect\citeauthoryear{{Soubiran} et~al.,}{{Soubiran}
  et~al.}{2018}]{soubiran2018}
{Soubiran} C.,  et~al., 2018, \mn@doi [\aap] {10.1051/0004-6361/201832795},
  \href {https://ui.adsabs.harvard.edu/abs/2018A&A...616A...7S} {616, A7}

\bibitem[\protect\citeauthoryear{{Tan} \& {Showman}}{{Tan} \&
  {Showman}}{2021}]{tan2021}
{Tan} X.,  {Showman} A.~P.,  2021, \mn@doi [\mnras] {10.1093/mnras/stab060},
  \href {https://ui.adsabs.harvard.edu/abs/2021MNRAS.502..678T} {502, 678}

\bibitem[\protect\citeauthoryear{{Tannock} et~al.,}{{Tannock}
  et~al.}{2021}]{tannock2021}
{Tannock} M.~E.,  et~al., 2021, \mn@doi [\aj] {10.3847/1538-3881/abeb67}, \href
  {https://ui.adsabs.harvard.edu/abs/2021AJ....161..224T} {161, 224}

\bibitem[\protect\citeauthoryear{{Tennyson} et~al.,}{{Tennyson}
  et~al.}{2016}]{tennyson2016}
{Tennyson} J.,  et~al., 2016, \mn@doi [Journal of Molecular Spectroscopy]
  {10.1016/j.jms.2016.05.002}, \href
  {https://ui.adsabs.harvard.edu/abs/2016JMoSp.327...73T} {327, 73}

\bibitem[\protect\citeauthoryear{{Tremblin}, {Amundsen}, {Chabrier}, {Baraffe},
  {Drummond}, {Hinkley}, {Mourier}  \& {Venot}}{{Tremblin}
  et~al.}{2016}]{tremblin2016}
{Tremblin} P.,  {Amundsen} D.~S.,  {Chabrier} G.,  {Baraffe} I.,  {Drummond}
  B.,  {Hinkley} S.,  {Mourier} P.,   {Venot} O.,  2016, \mn@doi [\apjl]
  {10.3847/2041-8205/817/2/L19}, \href
  {https://ui.adsabs.harvard.edu/abs/2016ApJ...817L..19T} {817, L19}

\bibitem[\protect\citeauthoryear{Triaud}{Triaud}{2018}]{Triaud_2018}
Triaud A. H. M.~J.,  2018, The Rossiter–McLaughlin Effect in Exoplanet
  Research.
Springer International Publishing, p. 1375–1401,
  \mn@doi{10.1007/978-3-319-55333-7_2}, \url
  {http://dx.doi.org/10.1007/978-3-319-55333-7_2}

\bibitem[\protect\citeauthoryear{{Visscher} \& {Moses}}{{Visscher} \&
  {Moses}}{2011}]{visscher2011}
{Visscher} C.,  {Moses} J.~I.,  2011, \mn@doi [\apj]
  {10.1088/0004-637X/738/1/72}, \href
  {https://ui.adsabs.harvard.edu/abs/2011ApJ...738...72V} {738, 72}

\bibitem[\protect\citeauthoryear{{Visscher}, {Lodders}  \& {Fegley}}{{Visscher}
  et~al.}{2010}]{visscher2010}
{Visscher} C.,  {Lodders} K.,   {Fegley} Bruce J.,  2010, \mn@doi [\apj]
  {10.1088/0004-637X/716/2/1060}, \href
  {https://ui.adsabs.harvard.edu/abs/2010ApJ...716.1060V} {716, 1060}

\bibitem[\protect\citeauthoryear{{Vos} et~al.,}{{Vos} et~al.}{2019}]{vos2019}
{Vos} J.~M.,  et~al., 2019, \mn@doi [\mnras] {10.1093/mnras/sty3123}, \href
  {https://ui.adsabs.harvard.edu/abs/2019MNRAS.483..480V} {483, 480}

\bibitem[\protect\citeauthoryear{{Vos}, {Faherty}, {Gagn{\'e}}, {Marley},
  {Metchev}, {Gizis}, {Rice}  \& {Cruz}}{{Vos} et~al.}{2022}]{vos2022}
{Vos} J.~M.,  {Faherty} J.~K.,  {Gagn{\'e}} J.,  {Marley} M.,  {Metchev} S.,
  {Gizis} J.,  {Rice} E.~L.,   {Cruz} K.,  2022, \mn@doi [\apj]
  {10.3847/1538-4357/ac4502}, \href
  {https://ui.adsabs.harvard.edu/abs/2022ApJ...924...68V} {924, 68}

\bibitem[\protect\citeauthoryear{{Vos} et~al.,}{{Vos} et~al.}{2023}]{vos2023}
{Vos} J.~M.,  et~al., 2023, \mn@doi [\apj] {10.3847/1538-4357/acab58}, \href
  {https://ui.adsabs.harvard.edu/abs/2023ApJ...944..138V} {944, 138}

\bibitem[\protect\citeauthoryear{{Wakeford} \& {Sing}}{{Wakeford} \&
  {Sing}}{2015}]{wakeford2015}
{Wakeford} H.~R.,  {Sing} D.~K.,  2015, \mn@doi [\aap]
  {10.1051/0004-6361/201424207}, \href
  {https://ui.adsabs.harvard.edu/abs/2015A&A...573A.122W} {573, A122}

\bibitem[\protect\citeauthoryear{{Wakeford}, {Visscher}, {Lewis}, {Kataria},
  {Marley}, {Fortney}  \& {Mandell}}{{Wakeford} et~al.}{2017}]{wakeford2017}
{Wakeford} H.~R.,  {Visscher} C.,  {Lewis} N.~K.,  {Kataria} T.,  {Marley}
  M.~S.,  {Fortney} J.~J.,   {Mandell} A.~M.,  2017, \mn@doi [\mnras]
  {10.1093/mnras/stw2639}, \href
  {https://ui.adsabs.harvard.edu/abs/2017MNRAS.464.4247W} {464, 4247}

\bibitem[\protect\citeauthoryear{{Welbanks}, {Madhusudhan}, {Allard}, {Hubeny},
  {Spiegelman}  \& {Leininger}}{{Welbanks} et~al.}{2019}]{welbanks2019}
{Welbanks} L.,  {Madhusudhan} N.,  {Allard} N.~F.,  {Hubeny} I.,  {Spiegelman}
  F.,   {Leininger} T.,  2019, \mn@doi [\apjl] {10.3847/2041-8213/ab5a89},
  \href {https://ui.adsabs.harvard.edu/abs/2019ApJ...887L..20W} {887, L20}

\bibitem[\protect\citeauthoryear{{Wilson}}{{Wilson}}{1999}]{wilson1999}
{Wilson} T.~L.,  1999, \mn@doi [Reports on Progress in Physics]
  {10.1088/0034-4885/62/2/002}, \href
  {https://ui.adsabs.harvard.edu/abs/1999RPPh...62..143W} {62, 143}

\bibitem[\protect\citeauthoryear{{Xuan} et~al.,}{{Xuan}
  et~al.}{2024a}]{xuan2024}
{Xuan} J.~W.,  et~al., 2024a, \mn@doi [arXiv e-prints]
  {10.48550/arXiv.2405.13128}, \href
  {https://ui.adsabs.harvard.edu/abs/2024arXiv240513128X} {p. arXiv:2405.13128}

\bibitem[\protect\citeauthoryear{{Xuan} et~al.,}{{Xuan}
  et~al.}{2024b}]{xuan2024_binary}
{Xuan} J.~W.,  et~al., 2024b, \mn@doi [\nat] {10.1038/s41586-024-08064-x},
  \href {https://ui.adsabs.harvard.edu/abs/2024Natur.634.1070X} {634, 1070}

\bibitem[\protect\citeauthoryear{{Xuan} et~al.,}{{Xuan}
  et~al.}{2024c}]{xuan2024_hip55507}
{Xuan} J.~W.,  et~al., 2024c, \mn@doi [\apj] {10.3847/1538-4357/ad1243}, \href
  {https://ui.adsabs.harvard.edu/abs/2024ApJ...962...10X} {962, 10}

\bibitem[\protect\citeauthoryear{{Zhang} et~al.,}{{Zhang}
  et~al.}{2021a}]{zhang2021}
{Zhang} Y.,  et~al., 2021a, \mn@doi [\nat] {10.1038/s41586-021-03616-x}, \href
  {https://ui.adsabs.harvard.edu/abs/2021Natur.595..370Z} {595, 370}

\bibitem[\protect\citeauthoryear{{Zhang}, {Snellen}  \& {Molli{\`e}re}}{{Zhang}
  et~al.}{2021b}]{zhang2021_bd}
{Zhang} Y.,  {Snellen} I. A.~G.,   {Molli{\`e}re} P.,  2021b, \mn@doi [\aap]
  {10.1051/0004-6361/202141502}, \href
  {https://ui.adsabs.harvard.edu/abs/2021A&A...656A..76Z} {656, A76}

\bibitem[\protect\citeauthoryear{{Zhang} et~al.,}{{Zhang}
  et~al.}{2024}]{zhang2024}
{Zhang} Y.,  et~al., 2024, \mn@doi [arXiv e-prints]
  {10.48550/arXiv.2409.16660}, \href
  {https://ui.adsabs.harvard.edu/abs/2024arXiv240916660Z} {p. arXiv:2409.16660}

\bibitem[\protect\citeauthoryear{Zhou et~al.,}{Zhou et~al.}{2019}]{Zhou_2019}
Zhou Y.,  et~al., 2019, \mn@doi [The Astronomical Journal]
  {10.3847/1538-3881/ab037f}, 157, 128

\bibitem[\protect\citeauthoryear{{Zhou}, {Bowler}, {Apai}, {Kataria}, {Morley},
  {Bryan}, {Skemer}  \& {Benneke}}{{Zhou} et~al.}{2022}]{zhou2022}
{Zhou} Y.,  {Bowler} B.~P.,  {Apai} D.,  {Kataria} T.,  {Morley} C.~V.,
  {Bryan} M.~L.,  {Skemer} A.~J.,   {Benneke} B.,  2022, \mn@doi [\aj]
  {10.3847/1538-3881/ac9905}, \href
  {https://ui.adsabs.harvard.edu/abs/2022AJ....164..239Z} {164, 239}

\bibitem[\protect\citeauthoryear{{de Regt} et~al.,}{{de Regt}
  et~al.}{2024}]{deregt2024}
{de Regt} S.,  et~al., 2024, \mn@doi [arXiv e-prints]
  {10.48550/arXiv.2405.10841}, \href
  {https://ui.adsabs.harvard.edu/abs/2024arXiv240510841D} {p. arXiv:2405.10841}

\makeatother
\end{thebibliography}




\appendix

\section{Best fit residuals for other nights} \label{sec:best_fit_appendix}

\begin{figure*}
	\includegraphics[width=\textwidth]{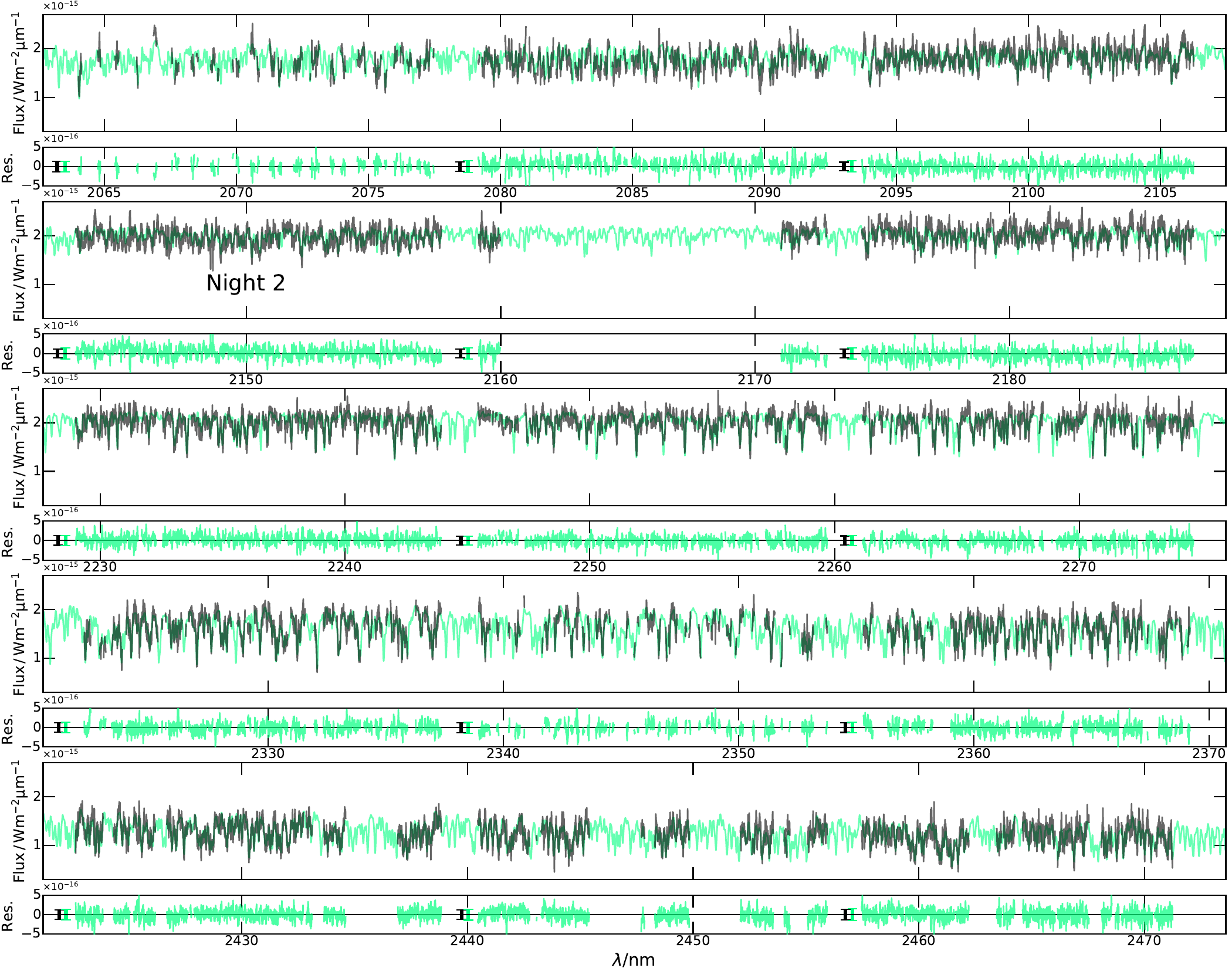}
    \caption{Best fit model from the retrieval of night 2 (green) against the observations (black). We also show the residuals between the model and the data in the bottom panels for each order. The mean photon noise for each order and detector is indicated with a black error bar, and the standard deviation of the residuals is shown in the green error bar.}
    \label{fig:best_fit_n2}
\end{figure*}

\begin{figure*}
	\includegraphics[width=\textwidth]{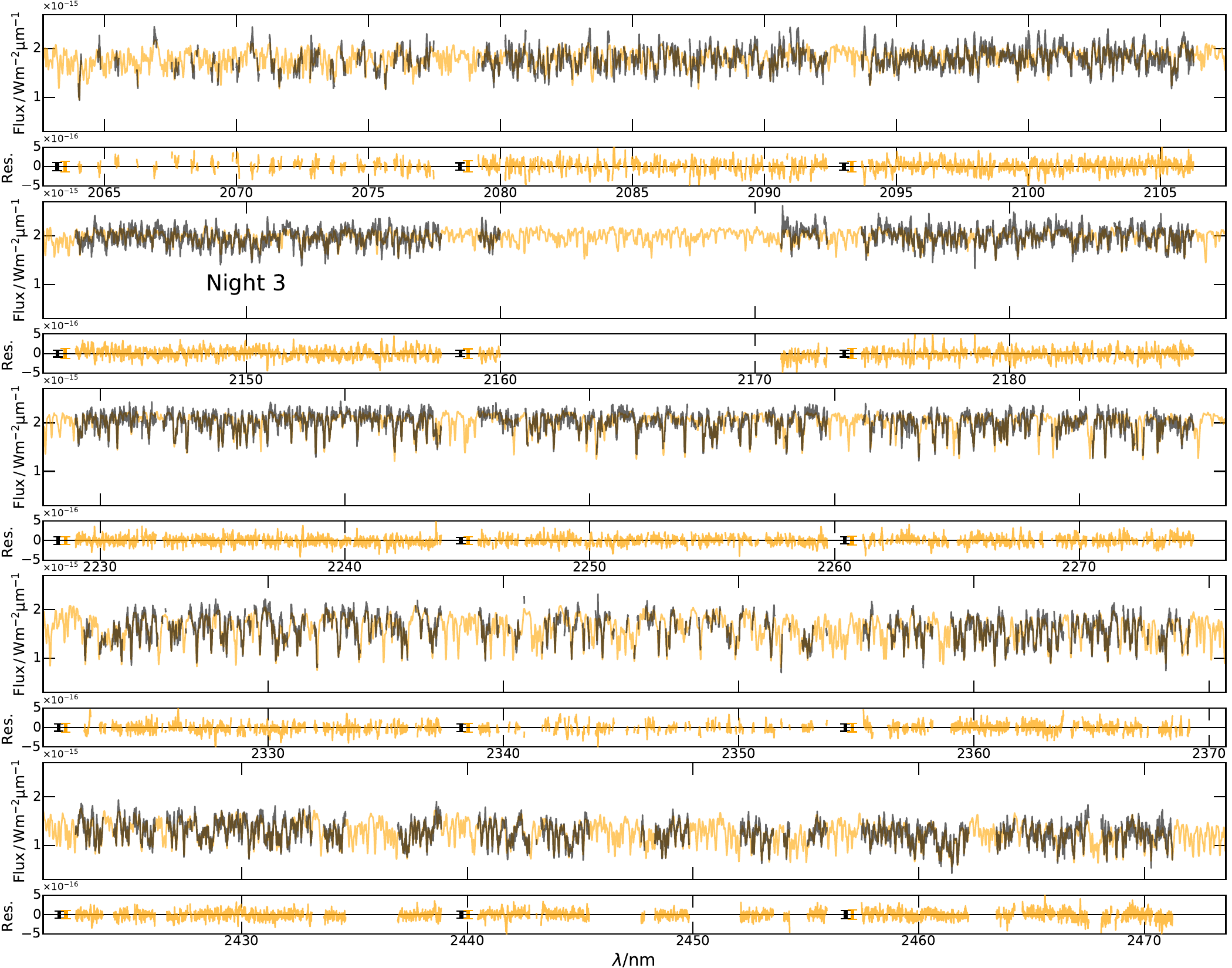}
    \caption{Best fit model from the retrieval of night 3 (orange) against the observations (black). We also show the residuals between the model and the data in the bottom panels for each order. The mean photon noise for each order and detector is indicated with a black error bar, and the standard deviation of the residuals is shown in the orange error bar.}
    \label{fig:best_fit_n3}
\end{figure*}

\begin{figure*}
	\includegraphics[width=\textwidth]{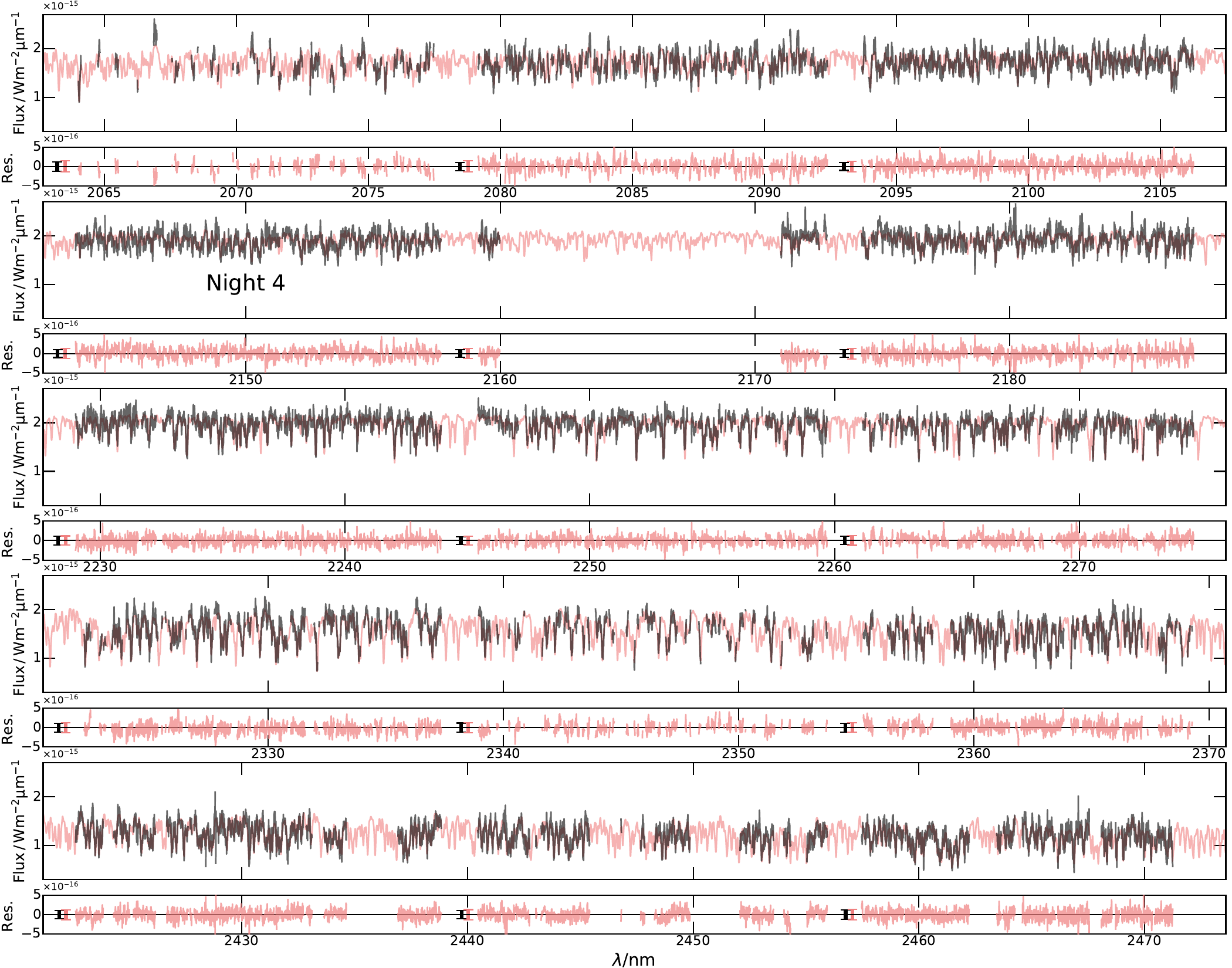}
    \caption{Best fit model from the retrieval of night 4 (red) against the observations (black). We also show the residuals between the model and the data in the bottom panels for each order. The mean photon noise for each order and detector is indicated with a black error bar, and the standard deviation of the residuals is shown in the red error bar.}
    \label{fig:best_fit_n4}
\end{figure*}


\bsp	
\label{lastpage}
\end{document}